\documentclass[aps,twocolumn,floatfix,amsmath,amsfonts,aps,longbibliography,groupedaddress]{revtex4-1}

\usepackage[utf8]{inputenc}

\usepackage{graphicx}
\usepackage{bm}
\usepackage{units}
\usepackage{color}
\usepackage{upgreek}
\usepackage{tikz}
\usepackage{amssymb, amsmath}
\usepackage{bbold}
\usepackage[normalem]{ulem}

\newlength{\hcolw}
\begin{document}

\title{Second Harmonic Generation of cuprous oxide in magnetic fields}
\author{Patric Rommel}
\author{J\"org Main}
\affiliation{Institut f\"ur Theoretische Physik 1, Universit\"at Stuttgart, 70550 Stuttgart, Germany}
\author{Andreas Farenbruch$^1$}
\author{Johannes Mund$^1$}
\author{Dietmar Fr\"ohlich$^1$}
\author{Dmitri R. Yakovlev$^{1,2}$}
\author{Manfred Bayer$^{1,2}$}
\author{Christoph Uihlein$^1$}
\affiliation{$^1$Experimentelle Physik 2, Technische Universit\"at Dortmund, 44221 Dortmund, Germany\\ $^2$Ioffe Institute, Russian Academy of Sciences, 194021 St.\ Petersburg, Russia}
\date{\today}

\begin{abstract}
\noindent Recently Second Harmonic Generation (SHG) for the yellow exciton series in cuprous oxide has been demonstrated [J. Mund \emph{et al.}, Phys. Rev. B \textbf{98}, 085203 (2018)].
Assuming perfect $O_{\mathrm{h}}$ symmetry, SHG is forbidden along certain high-symmetry axes. Perturbations can break this symmetry and forbidden transitions may become allowed.
We investigate theoretically the effect of external magnetic fields on the yellow exciton lines of cuprous oxide. We identify two mechanisms by which an applied magnetic field can
induce a second harmonic signal in a forbidden direction. First of all, a magnetic field by itself generally lifts the selection rules. In the Voigt configuration, an additional magneto-Stark
electric field appears. This also induces certain SHG processes differing from those induced by the magnetic field alone. Complementary to the manuscript by A. Farenbruch~\emph{et al.} [Phys.\ Rev.~B, submitted],
we perform a full numerical diagonalization of the exciton Hamiltonian including the complex valence band structure. Numerical results are compared with experimental data.
\end{abstract}

\maketitle

\section{Introduction}

Yellow excitons in cuprous oxide show a hydrogen-like series of peaks that has been followed up to a principal quantum number of $n=25$ by Kazimierczuk \emph{et al.}~\cite{GiantRydbergExcitons}.
Due to the influence of the crystal symmetry and complex valence band structure, the exciton spectrum shows typical deviations from a hydrogen spectrum. As the spherical symmetry is broken,
angular momentum is not a good quantum number anymore and, for example, splitting and mixing between P and different F states are observable~\cite{ObservationHighAngularMomentumExcitons}. Additionally, the symmetry
of the bands also significantly affects the selection rules for different optical processes~\cite{GiantRydbergExcitons,ImpactValence,frankmagnetoexcitonscuprousoxide} such as one-photon and two-photon excitation.

After the first theoretical treatment of two-photon processes in 1931~\cite{GoeppertMayer1931}, and their first experimental demonstration in the optical range in 1963~\cite{Hopfield1963},
nonlinear optical techniques have established themselves as useful methods for the study of electronic properties of solids~\cite{ShenNonlinearOptics,Froehlich1994}.
They complement linear tools due to different selection rules~\cite{Inoue1965}. For example, in one-photon absorption spectroscopy in
cuprous oxide the odd exciton states are excited, whereas in two-photon excitation, it is the even parity states.

One example of a nonlinear optical process is Second Harmonic Generation (SHG). In SHG, two incoming photons are combined into one outgoing photon
of double energy. Recently, Mund \emph{et al.}\ have demonstrated SHG for the yellow exciton series in cuprous oxide~\cite{Mund2018}. Here, the spectrum consists mainly of the even parity excitons.

The symmetry induced selection rules determine which exciton states can participate in SHG processes. Additional limitations concerning the polarization and direction
of the incoming and outgoing light exist. One important limitation is the existence of forbidden directions in the crystal, where SHG is not allowed due to symmetry reasons.
There is a number of ways in which a SHG signal can nevertheless be induced along such a direction~\cite{Saenger2006Orbital,Saenger2006DilMagSemic,Lafrantz2013MSE,Lafrentz2013SHGZnO,Brunne2015}.
In general, a perturbation can break the crystal symmetry and lift this selection rule. One possibility of such a perturbation is strain in the crystal. Even without the application
of an external strain, SHG has been observed for the yellow 1S orthoexciton in forbidden directions due to residual strain in the sample~\cite{Mund2019}. The excitons with higher principal quantum numbers
remain forbidden, since the energetic splitting due to the strain does not exceed their linewidths and the selection rule thus is not lifted for them~\cite{Mund2019}.

To observe the higher exciton states, a different method is required. In this work, we investigate the application of an external magnetic field.
For a discussion of the resulting SHG spectra, we have to differentiate two experimental geometries. In Faraday configuration, the magnetic field is applied parallel to the wave vector of the incident light, whereas in Voigt configuration
the two are perpendicular to each other. In the latter case, an additional term behaving like an effective electric field orthogonal to both the wave vector and the magnetic field
appears, breaking the inversion symmetry of the crystal. This leads to a mixing of odd and even parity excitons~\cite{Rommel2018} and thus to additional features in the SHG spectra.
In Faraday configuration this effective electric field is absent.

The induced SHG spectra significantly depend on the choice of polarization of the incoming and outgoing light. In particular, these dependencies differ among the mechanisms inducing SHG and
can therefore be used for their differentiation.

We focus on the diagonalization of the complete exciton Hamiltonian
including the valence band structure and on the detailed comparison of
numerical and experimental data for certain fixed choices of
polarization in this paper.
The polarization dependencies of the SHG spectra in general are
investigated more thoroughly in the manuscript by
Farenbruch~\emph{et al.}~\cite{Farenbruch2019},  {where SHG intensities are
treated as a function of the linear polarization angles of incoming
and outgoing light for certain fixed peaks. Additional mechanisms
for the production of SHG light beyond those in the present paper
are considered as well.}

In this manuscript, we first of all introduce the Hamiltonian of the exciton problem including the valence band structure of Cu$_2$O in Sec.~\ref{sec:TheoreticalFoundations}.
In Sec.~\ref{sec:Numerics}, we explain our numerical approach for the obtainment of the associated eigenvalues and eigenvectors. Following this, in Sec.~\ref{sec:SHG}
we show how these eigenvalues and eigenvectors can be used to simulate Second Harmonic Generation spectra and derive the selection rules in Sec.~\ref{sec:SelectionRules}.
We describe the experimental setup for SHG in Sec.~\ref{sec:Experiment}. In \ref{sec:Results}, the numerical results are shown 
and compared with experimental spectra. We conclude in Sec.~\ref{sec:Conclusion} and give a brief outlook.

\section{Theoretical Foundations \label{sec:TheoreticalFoundations}}

Excitons are bound states formed by an electron and a hole interacting via the Coulomb interaction. The Hamiltonian thus is generally given by~\cite{ImpactValence,frankmagnetoexcitonscuprousoxide}
\begin{equation}
 H = E_{\mathrm{g}} + H_{\mathrm{e}}\left(\boldsymbol{p}_{\mathrm{e}}\right)+H_{\mathrm{h}}\left(\boldsymbol{p}_\mathrm{h}\right) -\frac{e^2}{4\pi\varepsilon_0 \varepsilon \left|\boldsymbol{r}_{\mathrm{e}} - \boldsymbol{r}_{\mathrm{h}} \right|}\,,
 \label{eq:Hamiltonian}
\end{equation}
with the band gap energy $E_{\mathrm{g}}$, the dielectric constant $\varepsilon$, and the positions of the electron and hole $\boldsymbol{r}_\mathrm{e}$ and $\boldsymbol{r}_\mathrm{h}$.
The terms $H_{\mathrm{e}}$ and $H_{\mathrm{h}}$ denote the kinetic energy of the electron and hole, respectively.
Their form is determined by the symmetry of the bands and thus by the crystal structure. For Cu$_2$O, the lowest $\Gamma_6^+$ conduction band involved in the yellow and green series is parabolic and the kinetic energy of the electron is therefore given by
\begin{equation}
 H_{\mathrm{e}}\left(\boldsymbol{p}_{\mathrm{e}}\right)=\frac{\boldsymbol{p}_{\mathrm{e}}^{2}}{2m_{\mathrm{e}}}\,,
 \label{eq:ElectronKinetic}
\end{equation}
with the electron mass $m_\mathrm{e}$
The uppermost valence bands on the other hand are nonparabolic and involve correction terms of cubic symmetry. The Hamiltonian for the hole kinetic energy is~\cite{ImpactValence,Luttinger56CyclotronResonanceSemiconductors}
\begin{eqnarray}
H_{\mathrm{h}}\left(\boldsymbol{p}_{\mathrm{h}}\right) & = & H_{\mathrm{so}}+\left(1/2\hbar^{2}m_{0}\right)\left\{ \hbar^{2}\left(\gamma_{1}+4\gamma_{2}\right)\boldsymbol{p}_{\mathrm{h}}^{2}\right.\phantom{\frac{1}{1}}\nonumber \\
 & + & 2\left(\eta_{1}+2\eta_{2}\right)\boldsymbol{p}_{\mathrm{h}}^{2}\left(\boldsymbol{I}\cdot\boldsymbol{S}_{\mathrm{h}}\right)\phantom{\frac{1}{1}}\nonumber \\
 & - & 6\gamma_{2}\left(p_{\mathrm{h}1}^{2}\boldsymbol{I}_{1}^{2}+\mathrm{c.p.}\right)-12\eta_{2}\left(p_{\mathrm{h}1}^{2}\boldsymbol{I}_{1}\boldsymbol{S}_{\mathrm{h}1}+\mathrm{c.p.}\right)\phantom{\frac{1}{1}}\nonumber \\
 & - & 12\gamma_{3}\left(\left\{ p_{\mathrm{h}1},p_{\mathrm{h}2}\right\} \left\{ \boldsymbol{I}_{1},\boldsymbol{I}_{2}\right\} +\mathrm{c.p.}\right)\phantom{\frac{1}{1}}\nonumber \\
 & - & \left.12\eta_{3}\left(\left\{ p_{\mathrm{h}1},p_{\mathrm{h}2}\right\} \left(\boldsymbol{I}_{1}\boldsymbol{S}_{\mathrm{h}2}+\boldsymbol{I}_{2}\boldsymbol{S}_{\mathrm{h}1}\right)+\mathrm{c.p.}\right)\right\}. \phantom{\frac{1}{1}}
 \label{eq:HoleKinetic}
\end{eqnarray}
Here, $\{A,B\} = (AB + BA)/2$ is the symmetrized product and c.p.\ denotes cyclic permutation.
The quasi-spin $\boldsymbol{I}$ describes the degeneracy of the $\Gamma_5^+$ valence band Bloch functions. The spin-orbit coupling term
\begin{equation}
H_{\mathrm{so}}=\frac{2}{3}\Delta\left(1+\frac{1}{\hbar^{2}}\boldsymbol{I}\cdot\boldsymbol{S}_{\mathrm{h}}\right)
\end{equation}
couples the hole spin $\boldsymbol{S}_\mathrm{h}$ and the quasi-spin to the effective hole spin $\boldsymbol{J} = \boldsymbol{I} + \boldsymbol{S}_\mathrm{h}$.
This splits the valence band into the upper $\Gamma_7^+$ band with $J = 1/2$ and lower $\Gamma_8^+$ band with $J = 3/2$ by the spin-orbit splitting $\Delta$.

For a more comprehensive discussion we refer to Refs.~\cite{frankmagnetoexcitonscuprousoxide, franklinewidth, frankevenexcitonseries, Luttinger56CyclotronResonanceSemiconductors}.
The values of the material parameters for Cu$_2$O used in Eqs.~\eqref{eq:Hamiltonian}-\eqref{eq:CentorOfMass} are listed in Table~\ref{tab:Constants}.

\begin{table}

\protect\caption{Material parameters of Cu$_2$O used in Eqs.~\eqref{eq:Hamiltonian}-\eqref{eq:CentorOfMass}. }

\begin{centering}
\begin{tabular}{lll}
\hline 
band gap energy & $E_{\mathrm{g}}=2.17208\,\mathrm{eV}$ &~\cite{GiantRydbergExcitons}\tabularnewline
electron mass & $m_{\mathrm{e}}=0.99\, m_{0}$ &~\cite{HodbyEffectiveMasses}\tabularnewline
hole mass & $m_{\mathrm{h}} = 0.58\, m_0$ &~\cite{HodbyEffectiveMasses}\tabularnewline
dielectric constant & $\varepsilon=7.5$ &~\cite{LandoltBornstein1998DielectricConstant}\tabularnewline
spin-orbit coupling & $\Delta=0.131\,\mathrm{eV}$ &~\cite{SchoeneLuttinger}\tabularnewline
valence band parameters & $\gamma_{1}=1.76$ &~\cite{SchoeneLuttinger}\tabularnewline
 & $\gamma_{2}=0.7532$ &~\cite{SchoeneLuttinger}\tabularnewline
 & $\gamma_{3}=-0.3668$ &~\cite{SchoeneLuttinger}\tabularnewline
 & $\eta_{1}=-0.020$ &~\cite{SchoeneLuttinger}\tabularnewline 
 & $\eta_{2}=-0.0037$ &~\cite{SchoeneLuttinger}\tabularnewline
 & $\eta_{3}=-0.0337$ &~\cite{SchoeneLuttinger}\tabularnewline
fourth Luttinger parameter & $\kappa = -0.5$ &~\cite{frankmagnetoexcitonscuprousoxide}\tabularnewline
  $g$-factor of cond. band & $g_{\mathrm{c}}=2.1$ &~\cite{ArtyukhinGFactor}\tabularnewline
\hline
\label{tab:Constants}
\end{tabular}
\par\end{centering}

\end{table}

\subsection{Treatment of the magnetic field}
We are interested in calculating SHG spectra in an external magnetic field. We thus have to perform the 
minimal coupling $\boldsymbol{p}_{\mathrm{e}} \rightarrow \boldsymbol{p_{\mathrm{e}}} + e \boldsymbol{A}(\boldsymbol{r}_\mathrm{e})$ and
$\boldsymbol{p}_{\mathrm{h}} \rightarrow \boldsymbol{p}_{\mathrm{h}} - e \boldsymbol{A}(\boldsymbol{r}_\mathrm{h})$ with the vector
potential for a homogeneous field $\boldsymbol{A}(\boldsymbol{r}_{\mathrm{e,h}}) = \left(\boldsymbol{B} \times \boldsymbol{r}_{\mathrm{e,h}} \right)/2$. Additionally,
we include the interaction of the spins and the magnetic field~\cite{frankmagnetoexcitonscuprousoxide,Luttinger56CyclotronResonanceSemiconductors}
\begin{equation}
H_{B}=\mu_{\mathrm{B}}\left[g_{c}\boldsymbol{S}_{\mathrm{e}}+\left(3\kappa+g_{s}/2\right)\boldsymbol{I}-g_{s}\boldsymbol{S}_{\mathrm{h}}\right]\cdot\boldsymbol{B}/\hbar\,,
\end{equation}
with the Bohr magneton $\mu_{\mathrm{B}}$, the $g$-factor of the electron $g_c$ and the hole $g_s \approx 2$, and the fourth Luttinger parameter $\kappa$.

\subsection{Center-of-mass coordinates and magneto-Stark effect}
We perform our calculations in the center-of-mass reference frame given by the following transformation~\cite{Schmelcher1992,AVRON1978},
\begin{align}
 \boldsymbol{r} &= \boldsymbol{r}_{\mathrm{e}} - \boldsymbol{r}_{\mathrm{h}},\nonumber\\
 \boldsymbol{R} &= \frac{m_{\mathrm{e}}}{m_{\mathrm{e}} + m_{\mathrm{h}}} \boldsymbol{r}_{\mathrm{e}} + \frac{m_{\mathrm{h}}}{m_{\mathrm{e}} + m_{\mathrm{h}}} \boldsymbol{r}_{\mathrm{h}},\nonumber\\
 \boldsymbol{p} &= \hbar \boldsymbol{k} - \frac{e}{2} \boldsymbol{B} \times \boldsymbol{R} = \frac{m_h}{m_{\mathrm{e}} + m_{\mathrm{h}}} \boldsymbol{p}_{\mathrm{e}} - \frac{m_{\mathrm{e}}}{m_{\mathrm{e}} + m_{\mathrm{h}}} \boldsymbol{p}_{\mathrm{h}}, \nonumber\\
 \boldsymbol{P} &= \hbar \boldsymbol{K} + \frac{e}{2} \boldsymbol{B} \times \boldsymbol{r} = \boldsymbol{p}_{\mathrm{e}} + \boldsymbol{p}_{\mathrm{h}}\,,
 \label{eq:CentorOfMass}
\end{align}
with the hole mass $m_\mathrm{h}$.
In the center-of-mass reference frame,
$\boldsymbol{r}$ denotes the relative coordinate and $\boldsymbol{R}$ the position of the center of mass. The corresponding momenta are given by $\boldsymbol{p}$ and $\boldsymbol{P}$. The pseudomomentum $\boldsymbol{K}$ is
a constant of motion related to the center-of-mass momentum~\cite{AVRON1978}.
In general, the exciting laser will transfer a small but finite pseudomomentum $\hbar\boldsymbol{K}$ onto the exciton. Here, we make an approximation and only consider the leading term describing
the combined effect of the magnetic field and nonvanishing total momentum,
\begin{equation}
  H_{\mathrm{ms}} = \frac{\hbar e}{M} (\boldsymbol{K} \times \boldsymbol{B})\cdot \boldsymbol{r}\,,
  \label{eq:magneto-Stark-term}
\end{equation}
which is the magneto-Stark effect (MSE)~\cite{Lafrantz2013MSE}, with the total mass $M = m_\mathrm{e} + m_\mathrm{h}$. This term acts analogously to an additional effective electric field~\cite{Rommel2018}
\begin{equation}
  \boldsymbol{\mathcal{F}}_{\mathrm{mse}} = \frac{\hbar e}{M} (\boldsymbol{K} \times \boldsymbol{B})\,,
  \label{eq:EffectiveElectricField}
\end{equation}
perpendicular to both the direction of the incident light and the external magnetic field. The influence of this term will thus depend on the relative
orientations of the exciting laser and the magnetic field. In Faraday configuration, where $\boldsymbol{K}$ and $\boldsymbol{B}$ are parallel, it vanishes.
However, when $\boldsymbol{K}$ and $\boldsymbol{B}$ are chosen to be orthogonal to each other, this term is nonzero.

\subsection{Central-cell corrections}
Second Harmonic Generation principally involves the even parity states, like the S and D excitons. For a correct theoretical description, additional effects like the exchange interaction
and the Haken potential have to be included. The treatment for our numerical calculations will be as described in reference~\cite{frankevenexcitonseries} using the Haken potential.

\section{Numerical diagonalization of the exciton Hamiltonian}
\label{sec:Numerics}
To numerically calculate the eigenvalues and eigenstates of the exciton problem, we first express the stationary Schrödinger equation in a complete basis.
For the orbital angular part, we utilize the spherical harmonics with quantum numbers $L$ and $M$. Additional quantum numbers have to be introduced to
treat the quasi-spin $\boldsymbol{I}$ as well as the electron and hole spins $\boldsymbol{S}_\mathrm{e}$ and $\boldsymbol{S}_\mathrm{h}$. For our basis,
we first couple the hole spin and the quasi-spin to the effective hole spin $\boldsymbol{J} = \boldsymbol{I} + \boldsymbol{S}_\mathrm{h}$. Next we introduce the angular momentum
$\boldsymbol{F} = \boldsymbol{J} + \boldsymbol{L}$ and finally add the electron spin $\boldsymbol{S}_\mathrm{e}$ to get the total angular momentum $\boldsymbol{F}_t = \boldsymbol{F} + \boldsymbol{S}_\mathrm{e}$.
Note that the basis functions belonging to the quasi-spin $\boldsymbol{I}$ transform according to the irreducible representation $\Gamma_5^+$ in Cu$_2$O instead of the usual $\Gamma_4^+$ for a spin of unity. However, since
$\Gamma_5^+ = \Gamma_4^+ \otimes \Gamma_2^+$, we can perform the standard coupling of angular momenta and, to obtain the appropriate symmetry of the total state, multiply with $\Gamma_2^+$ at the end.
For the radial part we use the Coulomb-Sturmian functions~\cite{ImpactValence,SturmCommutationRecursion,CoulombSturmForNuclear}
\begin{equation}
 U_{N,L} = N_{N,L} \left(\frac{2r}{{b}}\right)^L \mathrm{e}^{-\frac{r}{{b}}} L^{2L+1}_N\left(\frac{2r}{{b}}\right)
 \label{eq:CoulombSturm}
\end{equation}
with the associated Laguerre polynomials $L^m_n(x)$ and a normalization factor $N_{N,L}$. Here, $N$ is the radial instead of the principal quantum number.
The parameter ${b}$ can in principle be freely chosen, but influences the convergence of the matrix diagonalization.
In total we thus get the basis states
\begin{equation}
 | \Pi \rangle = \left| N,\, L;\,\left(I,\, S_{\mathrm{h}}\right),\, J;\, F,\, S_{\mathrm{e}}; F_t,\, M_{F_t}\right\rangle \,,
 \label{eq:BasisStates}
\end{equation}
where we use $\Pi = \left\{ N,\, L;\,\left(I,\, S_{\mathrm{h}}\right),\, J;\, F,\, S_{\mathrm{e}}; F_t,\, M_{F_t} \right\}$ as an abbreviation for the set of used quantum numbers.
This basis has the advantage of being complete without the inclusion of the hydrogen continuum, but it is not orthogonal with respect to the standard scalar product.

Following Ref.~\cite{frankmagnetoexcitonscuprousoxide}, we express the Hamiltonian in spherical tensor notation. We will investigate spectra with
$\boldsymbol{B} \parallel$ $[001]$, $[110]$ and $[111]$. In each case, we will choose the quantization axis to be along the magnetic field and perform an according
rotation on the Hamiltonian. The expressions obtained for the Hamilton operator are found in the appendix of Ref.~\cite{frankmagnetoexcitonscuprousoxide}.
Using the ansatz
\begin{equation}
\left|\Psi\right\rangle = \sum_{\Pi} c_{\Pi} \left| \Pi \right\rangle
			\label{eq:Ansatz}
\end{equation}
for the exciton wave function $\left|\Psi\right\rangle$, the Schrödinger equation gets the form of a generalized eigenvalue problem,
\begin{equation}
 \boldsymbol{H} \boldsymbol{c} = E \boldsymbol{M} \boldsymbol{c}\,,
 \label{eq:GeneralizedEigenvalueEquation}
\end{equation}
with the Hamiltonian matrix $\boldsymbol{H}$ and the overlap matrix $\boldsymbol{M}$. We cut off our basis for appropriately large values of $N$ and $F$ to
obtain finite matrices. The solution is obtained by using a suitable Lapack routine~\cite{lapackuserguide3} and we thus get the eigenvalues $E$ and
the vector of coefficients $\boldsymbol{c}$ in the basis expression \eqref{eq:Ansatz}.

\section{Second Harmonic Generation}
\label{sec:SHG}
\begin{figure}
\includegraphics[width=\columnwidth]{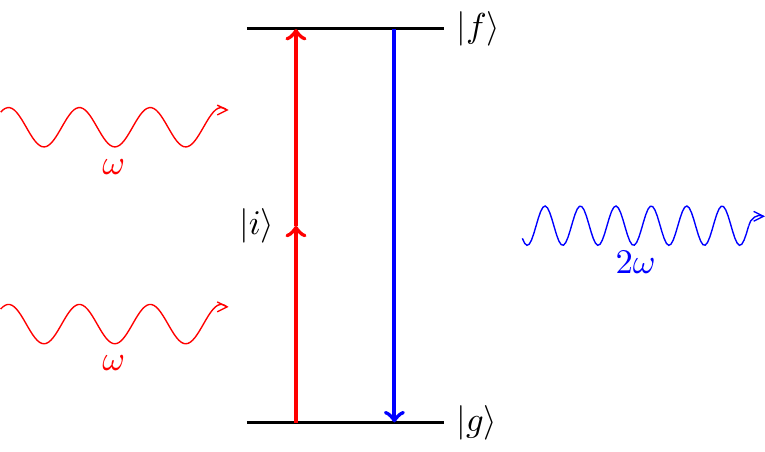}
\caption{Scheme of a Second Harmonic Generation process. The ground state of the crystal is denoted by $|g \rangle$, the resonantly stimulated exciton state by $|f \rangle$,
and the virtual intermediate state by $|i \rangle$.~\label{fig:SHGScheme}}
\end{figure}

Second Harmonic Generation is a process where two incoming photons are coherently transformed into one outgoing photon of doubled frequency as illustrated in Fig.~\ref{fig:SHGScheme}.
A given exciton state can only contribute to the SHG spectrum, if it is both two-photon and one-photon allowed. In the field-free case, only even exciton states can be excited in two-photon
transition processes. Since these are dipole forbidden, SHG can only be obtained by the addition of a quadrupole emission process. There are two conditions that determine the selection rules for these processes:
For the dominant contribution, the {envelope} wave function has to be nonvanishing at the origin~\cite{Inoue1965,frankjanpolariton}{, which requires an $L=0$ component,} and the exciton state has to have an admixture
of vanishing total spin $\boldsymbol{S} = \boldsymbol{S}_\mathrm{e} + \boldsymbol{S}_\mathrm{h} = 0${, since a spin flip is forbidden here}. Only the $\Gamma_5^+$ excitons of even parity fulfill both conditions.
{This can be seen by considering the resulting set of angular momenta. With $L=0$, $S = 0$ and $I = 1$, the rotational behavior for the exciton states is determined by the quasi-spin $\boldsymbol{I}$, which, as stated above, transforms according to $\Gamma_5^+$.}
We see that in the tensor product~{\cite{koster1963properties}}
\begin{equation}
 \Gamma_4^- \otimes \Gamma_4^- = \Gamma_1^+ \oplus \Gamma_3^+ \oplus \Gamma_4^+ \oplus \Gamma_5^+
\end{equation}
belonging to both the two-photon {($\Gamma_4^-$ for each incoming photon)} and quadrupole operator {($\Gamma_4^-$ for both the outgoing photon and the $\boldsymbol{K}$ vector)}, only the $\Gamma_5^+$ term contributes in leading order.

\subsection{Calculation of SHG intensities}
To simulate the SHG intensity spectra for a given polarization of the outgoing light $\boldsymbol{E}^\mathrm{out}$,
\begin{equation}
 I(2\omega) = \left|\sum_i E_i(2\omega) E^\mathrm{out}_i\right|^2
 \label{eq:SHGIntensity}
\end{equation}
with
\begin{equation}
  E_i(2\omega) \sim \sum_{j,k} \chi_{ijk}^{(2)} E_j(\omega) E_k(\omega)\,,
  \label{eq:SHGAmplitudesGeneral}
\end{equation}
we need to calculate the corresponding nonlinear susceptibilities
\begin{equation}
     \chi_{f,lmn}^{(2)} \sim \sum_i \frac{\langle g | V^Q_l |f \rangle \langle f|V^D_m|i \rangle \langle i|V^D_n|g\rangle}{(E_f - 2\hbar \omega - \mathrm{i}\Gamma_f)(E_i - \hbar \omega)}\,.
     \label{eq:nonlinearSusceptibility}
\end{equation}
Here, $D$ and $Q$ mark terms belonging to the excitation by two dipole steps and to the quadrupole emission process, respectively. The states involved are denoted by $|g\rangle$ for the ground
state of the crystal, $|f\rangle$ for the resonantly excited exciton state, and $|i\rangle$ for the virtual intermediate states.
The conditions of vanishing total spin (admixture of $\boldsymbol{S} = 0$) and nonvanishing wave function at the origin (admixture of $\boldsymbol{L} = 0$) imply that the strength of both processes is given by
the overlaps with the following states with irreducible representation $\Gamma_5^+$ \cite{frankjanpolariton}:
  \begin{align}
     |\pi_{yz}^Q\rangle &=\frac{1}{\sqrt{2}}\left(|1,-1\rangle_Q - |1,1\rangle_Q\right)\,, \nonumber\\
     |\pi_{xz}^Q\rangle &= \frac{\mathrm{i}}{\sqrt{2}}(|1,-1\rangle_Q + |1,1\rangle_Q)\,, \nonumber\\
     |\pi_{xy}^Q\rangle &=|1,0\rangle_Q\,, \phantom{ \frac{1}{\sqrt{2}}}
              \label{eq:quadrupoleOscillatorStates001}
  \end{align}
with
\begin{align}
 \left|F_t,M_{F_t}\right\rangle_Q &=\left|S,\,I;\,I+S,\,L;\,F_t,\,M_{F_t}\right\rangle \nonumber\\
                                  &=\left|0,\,1;\,1,\,0;\,1,\,M_{F_t}\right\rangle\,.
\end{align}
In Eq.~\eqref{eq:quadrupoleOscillatorStates001}, the quantization axis is chosen to be along the $[001]$ direction. If $[110]$ is chosen to be the $z$- and quantization axis, we instead have
  \begin{align}
     |\pi_{yz}^Q\rangle &= \frac{\mathrm{i}}{2}|1,-1\rangle_Q^{[110]}+ \frac{\mathrm{i}}{2}|1,1\rangle_Q^{[110]}+ \frac{1}{\sqrt{2}} |1,0\rangle_Q^{[110]}\,,\nonumber\\
     |\pi_{xz}^Q\rangle &= -\frac{\mathrm{i}}{2}|1,-1\rangle_Q^{[110]}- \frac{\mathrm{i}}{2}|1,1\rangle_Q^{[110]}+ \frac{1}{\sqrt{2}} |1,0\rangle_Q^{[110]}\,,\nonumber\\
     |\pi_{xy}^Q\rangle &= \frac{1}{\sqrt{2}} |1,-1\rangle_Q^{[110]}- \frac{1}{\sqrt{2}} |1,1\rangle_Q^{[110]}\,.
              \label{eq:quadrupoleOscillatorStates110}
  \end{align}
For the case of the quantization axis being parallel to the $[111]$ direction, we obtain
  \begin{align}
     |\pi_{yz}^Q\rangle &= \left(\frac{1}{2\sqrt{3}}-\frac{\mathrm{i}}{2}\right)|1,-1\rangle_Q^{[111]}+ \left(\frac{1}{2\sqrt{3}}+\frac{\mathrm{i}}{2}\right)|1,1\rangle_Q^{[111]}\nonumber\\
                        &\phantom{= } + \frac{1}{\sqrt{3}} |1,0\rangle_Q^{[111]}\,,\nonumber\\
     |\pi_{xz}^Q\rangle &= \left(\frac{1}{2\sqrt{3}}+\frac{\mathrm{i}}{2}\right)|1,-1\rangle_Q^{[111]}- \left(\frac{1}{2\sqrt{3}}-\frac{\mathrm{i}}{2}\right)|1,1\rangle_Q^{[111]}\nonumber\\
                        &\phantom{= } + \frac{1}{\sqrt{3}} |1,0\rangle_Q^{[111]}\,,\nonumber\\
     |\pi_{xy}^Q\rangle &= -\frac{1}{\sqrt{3}} |1,-1\rangle_Q^{[111]} + \frac{1}{\sqrt{3}} |1,1\rangle_Q^{[111]}\,.
              \label{eq:quadrupoleOscillatorStates111}
  \end{align}
For the two-photon excitation, we have to consider the coupling of $\Gamma_4^- \otimes \Gamma_4^- \rightarrow \Gamma_5^+$ for the two polarization vectors of the incoming light.
In this work, we only consider the case of two identical incoming photons with polarization $\boldsymbol{E}^{\mathrm{in}} = \left( E^{\mathrm{in}}_{1}\,, E^{\mathrm{in}}_{2}\,, E^{\mathrm{in}}_{3} \right)$.
The coupling coefficients as given in~\cite{koster1963properties} imply that the transition amplitudes for two-photon absorption with two dipole steps $\boldsymbol{O}_{\mathrm{TPDD}}$ can then by calculated using
the symmetrical cross product
\begin{align}
 \boldsymbol{O}_\mathrm{TPDD} \sim
 \boldsymbol{E}^\mathrm{in} \otimes \boldsymbol{E}^\mathrm{in} &= 
 \frac{1}{\sqrt{2}} \left(\begin{matrix}  E^{\mathrm{in}}_{2}E^{\mathrm{in}}_{3}+E^{\mathrm{in}}_{3}E^{\mathrm{in}}_{2} \\
 E^{\mathrm{in}}_{3}E^{\mathrm{in}}_{1}+E^{\mathrm{in}}_{1}E^{\mathrm{in}}_{3} \\ E^{\mathrm{in}}_{1}E^{\mathrm{in}}_{2}+E^{\mathrm{in}}_{2}E^{\mathrm{in}}_{1} \end{matrix} \right) \,,
 \label{eq:Two-Photon-Operator}
\end{align}
where the components give the amplitude for the excitation of a state transforming as $yz$ for $\boldsymbol{e}_1$, as $xz$ for $\boldsymbol{e}_2$ and as $xy$ for $\boldsymbol{e}_3$.
We see that, for example, light polarized along the $[110]$ direction will produce exciton states transforming according to the basis vector $xy$ of $\Gamma_5^+$.

For the quadrupole emission process, we similarly have to consider the coupling of the polarization vector $\boldsymbol{E}^\mathrm{out}$ of the outgoing light, determined by the analyzer in the
experiment, and the wave vector $\boldsymbol{K}$,
\begin{align}
 \boldsymbol{O}_\mathrm{Q} \sim \boldsymbol{K} \otimes \mathbf{E}^\mathrm{out} &= 
 \frac{1}{\sqrt{2}} \left(\begin{matrix}  K_{2}E^{\mathrm{out}}_{3}+E^{\mathrm{out}}_{3}K_{2} \\ K_{3}E^{\mathrm{out}}_{1}+E^{\mathrm{out}}_{1}K_{3} \\
 K_{1}E^{\mathrm{out}}_{2}+E^{\mathrm{out}}_{2}K_{1} \end{matrix} \right) \,.
 \label{eq:Quadrupole-Operator}
\end{align}
Analogously to the case of two-photon excitation, we can, for example, conclude that light polarized along the $[001]$ direction with a wave vector parallel to $[100]$ can
only be emitted by exciton states transforming as $xz$.

\subsection{Dipole emission process}
In Voigt configuration, considered in some of the spectra here, an effective electric field arises. This electric field breaks the inversion symmetry of the crystal and
mixes states of different parity. This will also make certain SHG processes involving a dipole emission step allowed.
Similar to the case of the two-photon excitation and quadrupole emission processes, the strength of these dipole emission processes are given by the overlaps
with the three states of symmetry $\Gamma_4^-$ as derived in Refs.~\cite{ImpactValence,frankmagnetoexcitonscuprousoxide,frankjanpolariton}:
		                         \begin{align}
                                \left|\pi_x^D\right\rangle & = \frac{\mathrm{i}}{\sqrt{2}}\left(\left|2,-1\right\rangle_D+\left|2,1\right\rangle_D\right)\,,\nonumber\\
                                \left|\pi_y^D\right\rangle & = \frac{1}{\sqrt{2}}\left(\left|2,-1\right\rangle_D-\left|2,1\right\rangle_D\right)\,,\nonumber\\
                                \left|\pi_z^D\right\rangle & = \frac{\mathrm{i}}{\sqrt{2}}\left(\left|2,-2\right\rangle_D-\left|2,2\right\rangle_D\right)\,,
                                         \label{eq:dipolOscillatorStates001}
		                         \end{align}
		                         with
   \begin{align}
		                         \left|F_t,M_{F_t}\right\rangle_D &= \left|S,\,I;\,I+S,\,L;\,F_t,\,M_{F_t}\right\rangle \nonumber\\
		                                                           &= \left|0,\,1;\,1,\,1;\,F_t,\,M_{F_t}\right\rangle\,.
   \end{align}
Again, Eq.~\eqref{eq:dipolOscillatorStates001} gives the result with quantization axis along $[001]$. For $[110]$, we get
   \begin{align}
                                \left|\pi_x^D\right\rangle & = -\frac{\mathrm{i}}{2} |2,-2\rangle_D^{[110]} + \frac{1}{2} |2,-1\rangle_D^{[110]}\nonumber\\
                                                           & \phantom{= } - \frac{1}{2} |2,1\rangle_D^{[110]} + \frac{\mathrm{i}}{2} |2,2\rangle_D^{[110]}\,, \nonumber\\
                                \left|\pi_y^D\right\rangle & = \frac{\mathrm{i}}{2} |2,-2\rangle_D^{[110]} + \frac{1}{2} |2,-1\rangle_D^{[110]}\nonumber\\
                                                           & \phantom{= } - \frac{1}{2} |2,1\rangle_D^{[110]} - \frac{\mathrm{i}}{2} |2,2\rangle_D^{[110]}\,,\nonumber\\
                                \left|\pi_z^D\right\rangle & = - \frac{1}{\sqrt{8}} |2,-2\rangle_D^{[110]} + \frac{\sqrt{3}}{2}|2,0\rangle_D^{[110]} \nonumber\\
                                                           & \phantom{= } + \frac{1}{\sqrt{8}} |2,2\rangle_D^{[110]} \,,\label{eq:dipolOscillatorStates110}                                         
   \end{align}
and for the $[111]$ direction
   \begin{align}
                                \left|\pi_x^D\right\rangle & =\left(\frac{1}{\sqrt{18}} -\frac{\mathrm{i}}{6} \right) |2,-2\rangle_D^{[111]} +\left(\frac{1}{6} + \frac{\mathrm{i}}{\sqrt{12}} \right) |2,-1\rangle_D^{[111]} \nonumber\\
                                                           & - \left(\frac{1}{6} + \frac{\mathrm{i}}{\sqrt{12}} \right) |2,1\rangle_D^{[111]} + \left(\frac{1}{\sqrt{18}} + \frac{\mathrm{i}}{6} \right) |2,2\rangle_D^{[111]}\nonumber \\
                                                           & + \frac{1}{3}|2,0\rangle_D^{[111]}\,,\nonumber\\
                                \left|\pi_y^D\right\rangle & =\left(\frac{1}{6} - \frac{\mathrm{i}}{\sqrt{12}} \right) |2,1\rangle_D^{[111]} + \left(\frac{1}{\sqrt{18}} - \frac{\mathrm{i}}{6} \right) |2,2\rangle_D^{[111]}\nonumber\\ 
                                                           & + \left(\frac{1}{6} - \frac{\mathrm{i}}{\sqrt{12}} \right) |2,-1\rangle_D^{[111]} + \left(\frac{1}{\sqrt{18}} +\frac{\mathrm{i}}{6} \right) |2,-2\rangle_D^{[111]}\nonumber\\
                                                           & +  \frac{1}{3}|2,0\rangle_D^{[111]}\,,\nonumber \\
                                \left|\pi_z^D\right\rangle & = \frac{1}{\sqrt{3}}|2,0\rangle_D^{[111]} + \frac{1}{3} |2,1\rangle_D^{[111]} - \frac{2}{\sqrt{18}} |2,2\rangle_D^{[111]}\nonumber\\
                                                           & - \frac{2}{\sqrt{18}} |2,-2\rangle_D^{[111]} - \frac{1}{3} |2,-1\rangle_D^{[111]} \,.\label{eq:dipolOscillatorStates111}
   \end{align}

\subsection{Linewidths}
In addition to the transition matrix elements discussed above, the nonlinear susceptibilities \eqref{eq:nonlinearSusceptibility} also depend
on the linewidths of the involved exciton states $\Gamma_f$. 
The homogeneous linewidths of the involved states are for the most part unknown. Additionally,
the strong mixing of states makes accurate assignments of states difficult. Various attempts to incorporate the linewidths in a more
detailed way did not lead to results in better agreement with experiment than the simple assumption of a constant
linewidth of $\Gamma = 150 \, \mathrm{\mu eV}$ for all states. This linewidth also approximately reproduces the widths of the dominant S and D states, as visible in
Figs.~\ref{fig:DetermineA} and \ref{fig:DetermineA04}.
We will thus use this simple approach for our numerical calculations.

\subsection{Relative strength of dipole and quadrupole emission processes}
\label{subsec:RelStrength}
\begin{figure}
\includegraphics[width=\columnwidth]{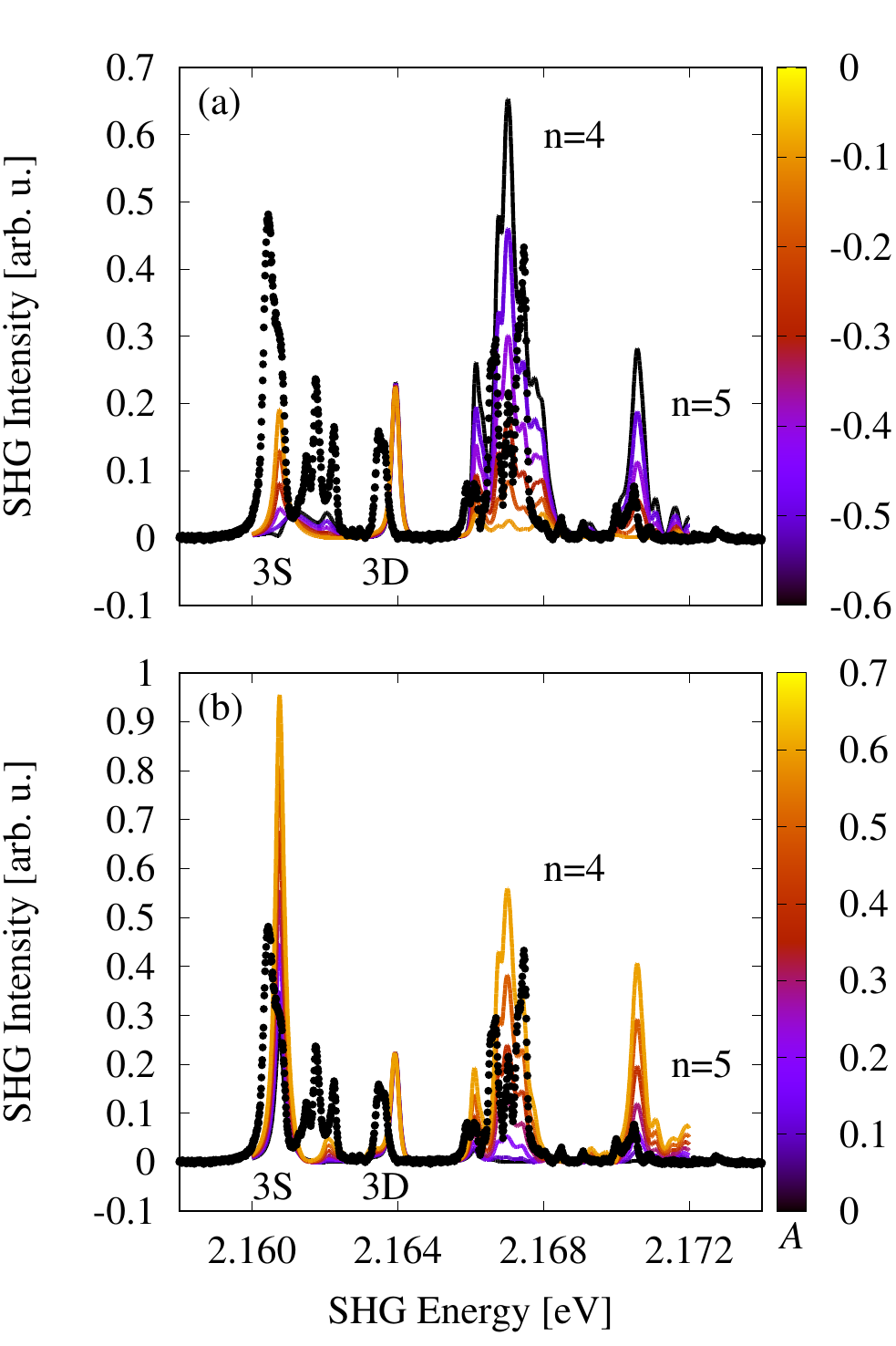}
 \caption{Comparison of experimental (black) and numerical (color palette) SHG spectra in arbitrary units with
 $\boldsymbol{E}^{\mathrm{in},2}~\parallel~[1 1 0]$, $\boldsymbol{E}^{\mathrm{out},2}~\parallel~[0 0 1]$ for different values of $A$ as defined in Eq.~\eqref{eq:DefinitionA}.
 The wave vector points along the $[1\overline{1}0]$ axis
 and the magnetic field is applied
 in Voigt geometry in $[110]$ direction and has a strength of $B~=~6\,\mathrm{T}$. The main features shown belong to principal quantum numbers $n = 3$ and $n=4$.
 The color encodes the value of $A$, which parametrizes the relative strength of dipole emission processes to quadrupole ones.
 We show the comparison for (a) negative and (b) positive values of $A$.
 }
 \label{fig:DetermineA}
\end{figure}

\begin{figure}
 \includegraphics[width=\columnwidth]{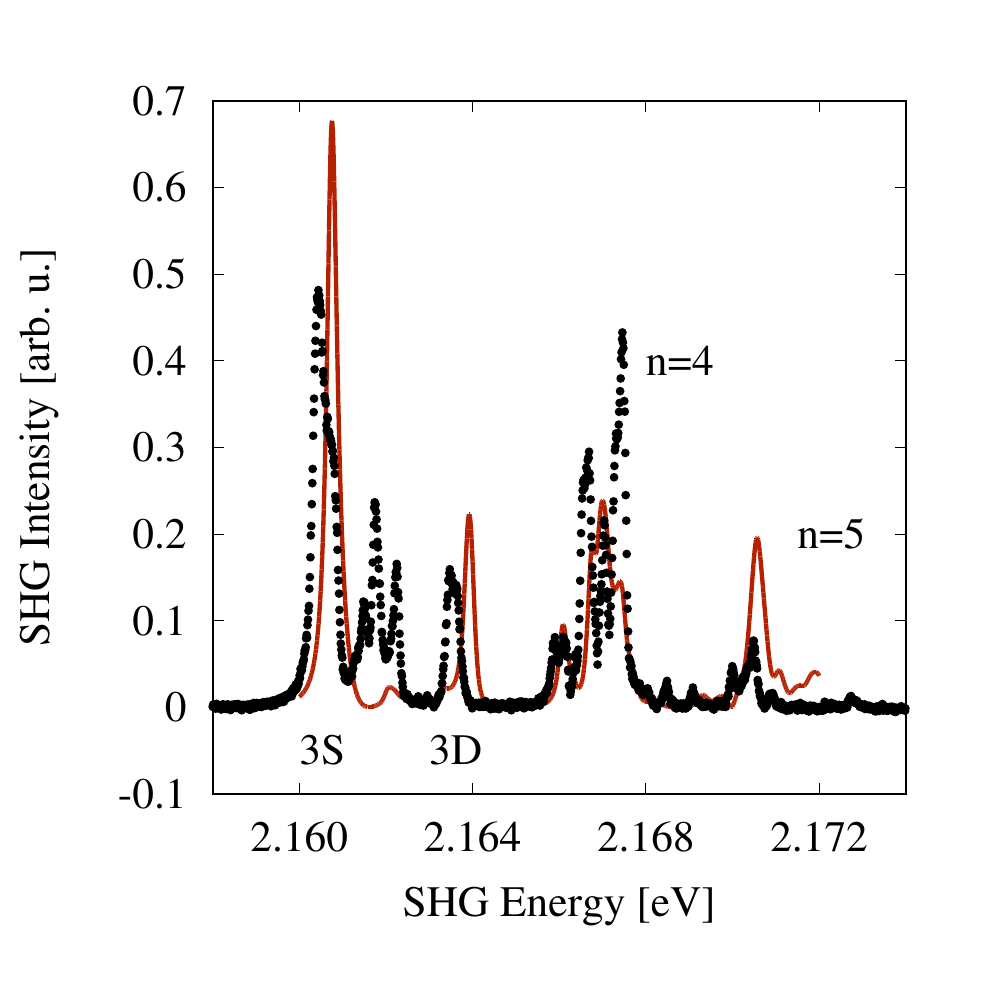}\caption{Same as Fig.~\ref{fig:DetermineA} but with a fixed value of $A=0.4$.
 With this value, the numerical spectrum matches the experimental spectrum reasonably well. {Choosing a higher value for $A$ leads to the 3S peak being excessively high, whereas a lower value results in a too weak $n=4$ manifold.} For further discussion, see text.
 \label{fig:DetermineA04}
 }
\end{figure}
In spectra where both quadrupole and dipole emission processes play a role, their relative oscillator strengths have to be considered. According
to Ref.~\cite{frankjanpolariton}, the combined transition matrix elements for both processes is given by
\begin{align}
 M \sim \lim_{r\rightarrow 0} \bigg[-\mathrm{i}\left(\tilde{M}_v^*+\tilde{M}_c^*\right)\frac{\partial}{\partial r}\langle T^D_{\mathrm{out}}|\Psi\rangle \nonumber\\
 +\left(-(1-\alpha)\tilde{M}_v^*+\alpha\tilde{M}_c^*\right)\frac{K}{\sqrt{6}}\langle T^Q_{\mathrm{out}}|\Psi\rangle \bigg] \,,
\label{eq:totalMatrixElement}
\end{align}
with the exciton wave function $|\Psi\rangle$. The parameter $\alpha$ relates to the chosen center-of-mass transformation by $\alpha = m_\mathrm{e}/(m_\mathrm{e}+ m_\mathrm{h}) {= 0.63}$.
The states $|T^D_{\mathrm{out}}\rangle$ and $|T^Q_{\mathrm{out}}\rangle$ are related to the states in Eqs.~\eqref{eq:quadrupoleOscillatorStates001} and \eqref{eq:dipolOscillatorStates001}
via
\begin{align}
|T^D_{\mathrm{out}}\rangle &= \hspace{.5em} \sum_{i \in \{x,y,z\} } E^\mathrm{out}_i | \pi_i^D \rangle \,, \\
|T^Q_{\mathrm{out}}\rangle &= \sum_{v \in \{ yz,xz,xy\} } \sqrt{2} (\boldsymbol{E}^\mathrm{out} \otimes \hat{\boldsymbol{K}})_v | \pi_v^Q \rangle \,,
\end{align}
with the normalized wave vector $\hat{\boldsymbol{K}}$. We see that the correct calculation of the SHG intensities requires the values for the constants $\tilde{M}_v^{{*}}$ and $\tilde{M}_c^{{*}}$.
These are independent of the exciton state and the magnetic field.
We rescale and rewrite Eq.~\eqref{eq:totalMatrixElement} as
\begin{equation}
 M \sim \lim_{r\rightarrow 0} \bigg[-\mathrm{i}A\frac{\partial}{\partial r}\langle T^D_{\mathrm{out}}|\Psi\rangle +\frac{K}{\sqrt{6}}\langle T^Q_{\mathrm{out}}|\Psi\rangle \bigg] \,,
\end{equation}
where 
\begin{equation}
 A = \frac{\tilde{M}_v^*+\tilde{M}_c^*}{-(1-\alpha)\tilde{M}_v^*+\alpha\tilde{M}_c^*}\,.
 \label{eq:DefinitionA}
\end{equation}
$A$ now parametrizes the relative contribution of dipole and quadrupole emission processes, i.e., for $|A|~\rightarrow~\infty$ the spectrum is only determined by dipole processes, whereas for $A~\rightarrow~0$, they play no role.
In Fig.~\ref{fig:DetermineA} we show a comparison of experimental and numerical spectra for a particular
strength of the magnetic field $B = 6\,\mathrm{T}$. Since the SHG spectrum is sensitive to the relative contributions of the quadrupole and dipole emission processes, we can use this comparison to estimate the value of $A$.
Reasonable agreement is achieved for $A = 0.4$, see Fig.~\ref{fig:DetermineA04}, and we will choose this value for $A$ for our further calculations.
{This allows us to estimate the ratio of $\tilde{M}_c^*$ to $\tilde{M}_v^*$. Using Eq.~\eqref{eq:DefinitionA}, we find
\begin{equation}
  \frac{\tilde{M}_c^*}{\tilde{M}_v^*} = - \frac{(1-\alpha)A+1}{1-\alpha A} \approx -1.5 \,.
\end{equation}
Note that this result can only be taken as a rough estimate. We chose the value of $A$ mainly on the basis of the agreement with the 3S and 3D states and the $n=4$ manifold. Still, the accordance between experiment and theory is not perfect, especially for the lines between the 3S and 3D states, which are not reproduced very well in the simulations. Presumably, this is due to the simplified treatment of the linewidths.
}
We {also} see that the feature around the 3S states comes out too strong. This is also observed in some of the following spectra. Two remarks are important here. First of all, the linewidth of the 3P state is
around $500\,\mu\mathrm{eV}$~\cite{GiantRydbergExcitons} and thus considerably larger than the value used here. Broader lines generally have weaker SHG intensities, exceptions may be caused by interference between different states.
The second remark concerns the line positions of the even exciton states being influenced by the central-cell corrections. Since the central-cell corrections are only an approximation, the positions of the even excitons
are not reproduced as faithfully  as the positions of the odd states. Instead, the numerical S and D excitons are shifted to slightly higher energies as compared to experiment, an effect also observed in Refs.~\cite{frankevenexcitonseries,Rommel2018}.
The reduced energetic distance between the S and P states probably leads to a stronger mixing and thus, for SHG with a dipole emission step, to an overestimated intensity. The reverse will hold for the D states.

\section{Discussion of selection rules}
\label{sec:SelectionRules}
Second Harmonic Generation is principally forbidden in inversion symmetric crystals such as Cu$_2$O. To see this, we consider the SHG amplitude $E_i(2\omega)$ given in Eq.~\eqref{eq:SHGAmplitudesGeneral}.
The application of the inversion operation switches the signs
of the amplitudes $E$, but leaves the susceptibility $\chi$ invariant due to the symmetry of the crystal. It follows that the amplitudes $E_i(2\omega)$ vanish unless the inversion symmetry is broken.

\subsection{Quadrupole and electric-field induced dipole emission}
\label{subsec:QuadEDipole}
A two-photon absorption process can only
excite even parity states. For two incoming photons with identical polarization $\boldsymbol{E}^\mathrm{in}$, the corresponding two-photon absorption amplitudes are given by the symmetrical cross product~\eqref{eq:Two-Photon-Operator},
\begin{equation}
 \boldsymbol{O}_\mathrm{TPDD} \sim  \boldsymbol{E}^\mathrm{in} \otimes \boldsymbol{E}^\mathrm{in}\,.
\end{equation}
The stimulated excitons transform according to $\Gamma_5^+$. Due to their parity, they cannot emit photons in a dipole process.
For SHG to become possible, a perturbation has to break the inversion symmetry. In the field-free case, this is accomplished by the wave vector $\boldsymbol{K}$, allowing for quadrupole emission processes.
For a given polarization of outgoing SHG light $\boldsymbol{E}^\mathrm{out}$, the associated quadrupole transition amplitudes transform like the symmetrical cross product~\eqref{eq:Quadrupole-Operator},
\begin{equation}
 \boldsymbol{O}_\mathrm{Q} \sim \boldsymbol{K} \otimes \boldsymbol{E}^\mathrm{out}\,.
 \label{eq:Quadrupole-Operator2}
\end{equation}
Combining both steps, the wave-vector induced SHG amplitude $M_K$ is proportional to
\begin{equation}
 M_K(\boldsymbol{E}^\mathrm{in},\boldsymbol{E}^\mathrm{out},\boldsymbol{K}) \sim (\boldsymbol{E}^\mathrm{in} \otimes \boldsymbol{E}^\mathrm{in})\cdot(\boldsymbol{K} \otimes \boldsymbol{E}^\mathrm{out})\,.
 \label{eq:SHGAmplitudeK}
\end{equation}

A different way to break the inversion symmetry is to apply an external electric field.
This causes the $\Gamma_5^+$ excitons to gain an admixture of dipole-allowed $\Gamma_4^-$ states. This makes a dipole emission step possible.
The first order transition amplitude
for a dipole emission process from the $\Gamma_5^+$ exciton state $\psi^{5+}_i$ to the ground state of the crystal $|g\rangle$ is then given by
\begin{equation}
 O^\mathcal{F}_{\mathrm{D},i} \sim \sum_{j,k,l} E^\mathrm{out}_l \mathcal{F}_j \frac{\langle \psi^{5+}_i | V^\mathcal{F}_j | \psi^{4-}_k \rangle \langle \psi^{4-}_k | V^D_l | g \rangle }{E_k - \hbar \omega_\mathrm{out}}\,.
\end{equation}
Here, $V^D_l$ is the term of the dipole operator belonging to the component of the polarization of the outgoing light $E^\mathrm{out}_l$ and transforms according to $\Gamma_4^-$, as does the perturbation $V^\mathcal{F}_j$
belonging to the component $\mathcal{F}_j$ of the electric field.
The projection operator
\begin{equation}
  P^{4-} = \sum_k \frac{ | \psi^{4-}_k \rangle \langle \psi^{4-}_k |}{E_k - \hbar \omega_\mathrm{out}}
\end{equation}
transforms according to the irreducible representation $\Gamma_1^+$.
For a nonvanishing contribution, the total matrix element for a given term in the sum over $j$ and $l$ has to transform as $\Gamma_1^+$. Since the ground state $|g\rangle$ belongs to $\Gamma_1^+$, this can only happen
if the complete operator between bra $\langle \psi^{5+}_i |$ and ket $|g\rangle$ transforms as $\Gamma_5^+$.
The matrix element is thus proportional to the group theoretical coupling coefficients belonging to the product $\Gamma_4^- \otimes \Gamma_4^- \rightarrow \Gamma_5^+$, which give
the symmetrical cross product of the outgoing polarization $\boldsymbol{E}^\mathrm{out}$ with the electric field $\boldsymbol{\mathcal{F}}$,
\begin{equation}
 \boldsymbol{O}^\mathcal{F}_\mathrm{D} \sim {\boldsymbol{\mathcal{F}}} \otimes \boldsymbol{E}^\mathrm{out}\,.
 \label{eq:FDipole}
\end{equation}
Taking the two-photon absorption step into account, the electric-field induced SHG amplitude $M_\mathcal{F}$ is given by
\begin{equation}
 M_\mathcal{F}( \boldsymbol{E}^\mathrm{in}, \boldsymbol{E}^\mathrm{out}, \boldsymbol{\mathcal{F}}) \sim (\boldsymbol{E}^\mathrm{in} \otimes \boldsymbol{E}^\mathrm{in}) \cdot ({\boldsymbol{\mathcal{F}}} \otimes \boldsymbol{E}^\mathrm{out})\,.
                                                                                       \label{eq:SHGAmplitudeF}
\end{equation}
Comparing formulas \eqref{eq:SHGAmplitudeK} and \eqref{eq:SHGAmplitudeF}, we see the close analogy between the wave vector $\boldsymbol{K}$ and the electric field $\boldsymbol{\mathcal{F}}$ in inducing a SHG signal.

\subsection{Separating magneto-Stark effect and Zeeman effect in forbidden directions}
Second Harmonic Generation induced by the finite wave vector $\boldsymbol{K}$ is not always possible. If $\boldsymbol{K}$ is directed along an axis with a $C_2$ symmetry, an argument analogous to the one
for the inversion symmetry above shows that the SHG signal vanishes. Group theoretically, the two-photon absorption process can only excite longitudinal states belonging to the irreducible representation $\Gamma_1$ in $C_2$.
Only transversal states of symmetry $\Gamma_2$ can emit a photon. The crystal has a $C_2$ symmetry for rotations around
the $[001]$ and $[110]$ axis and their equivalents. SHG is thus forbidden along those directions.

The direction investigated in this paper is given by $\boldsymbol{K}$ $\parallel$ $[1\overline{1}0]$.
To produce a SHG signal, the $C_2$ symmetry has to be broken and states belonging to $\Gamma_1$ and $\Gamma_2$ have to be coupled to each other. To this end, we consider
the application of an external magnetic field. In Faraday configuration, the $C_2$ symmetry remains. It is therefore necessary to apply the field in Voigt configuration.
We choose $\boldsymbol{B}$ $\parallel$ $[110]$.
In this case, in addition to the magnetic field the magneto-Stark electric field has to be treated as well. According to Eq.~\eqref{eq:EffectiveElectricField} it is directed along
$\boldsymbol{\mathcal{F}}$ $\parallel$ $[001]$.
Both the magnetic field and the electric field each induce a contribution to
the SHG signal. The magnetic field breaks the $C_2$ symmetry and produces exciton eigenstates containing $\Gamma_1$ and $\Gamma_2$ admixtures as necessary. The emission step
still results from a quadrupole process and can therefore be described using Eq.~\eqref{eq:Quadrupole-Operator2}.
For the electric field, the description given in Sec.~\ref{subsec:QuadEDipole} is valid and Eq.~\eqref{eq:SHGAmplitudeF} can be used if the Zeeman splitting is weak.

Evaluating these formulas in the given configuration reveals that the quadrupole emission induced by the Zeeman effect and the dipole emission induced by the magneto-Stark effect
have orthogonal polarizations to each other. Orienting the analyzer according to $\boldsymbol{E}^{\mathrm{out},1}$~$\parallel$~$[ 1 1 0 ]$, only electric-field induced dipole processes are possible.
By contrast, for $\boldsymbol{E}^{\mathrm{out},2}$ $\parallel$ $[ 0 0 1 ]$ only quadrupole emission is observable. This allows for the possibility of separating
Zeeman-induced SHG from magneto-Stark-induced SHG. Combining $\boldsymbol{E}^{\mathrm{out},1}$ with $\boldsymbol{E}^{\mathrm{in},1}$ $\parallel$ $[ 1 1 \sqrt{2} ]$, a SHG signal caused only by the electric field can be observed.
To accomplish the same for the Zeeman effect, we need to understand the effect of the magnetic field in greater detail.

\subsection{Symmetry reduction by the magnetic field}
\label{subsec:SymmetryMagnetic}
A magnetic field reduces the symmetry of the system and leads to a mixing of previously uncoupled states.
The principal effect relevant for SHG production is the coupling of states in the degenerate spaces belonging to the irreducible representation $\Gamma_5^+$ and the consequent lifting of their degeneracy.
As the magnetic field is of even parity, SHG is only produced by the combination of a two-photon excitation with a quadrupole emission process involving these states. 
Using Eqs.~\eqref{eq:Two-Photon-Operator} and \eqref{eq:Quadrupole-Operator}, a sum of basis vectors transforming like the $\Gamma_5^+$ states $\psi^{5+}_{yz}$, $\psi^{5+}_{xz}$ and $\psi^{5+}_{xy}$ can be assigned to
the two-photon and quadrupole amplitudes for a given pair of polarizations of the incoming and outgoing light. An exciton state can generally be excited in a two-photon absorption process
if it has a nonzero overlap with the resulting vector for the two-photon amplitudes. It can emit in a quadrupole step if it has a nonzero overlap with the resulting vector for the quadrupole amplitudes.
SHG is thus possible if the admixture by the magnetic field produces exciton states fulfilling both conditions.

To apply these rules in specific cases, we first need to understand the effect of the magnetic field on the exciton states.
To this end we will use a perturbation theoretical approach, considering the mixture of the $\Gamma_5^+$ states to leading order in $\boldsymbol{B}$.
We have to consider the lifting of the degeneracy through the magnetic field, leading to mixtures of zeroth order when the splitting is larger than the linewidths of the states.
Using the coupling coefficients in Ref.~\cite{koster1963properties}, we see that we have to diagonalize the following matrix with the identification $\boldsymbol{e}_1 = \psi^{5+}_{yz}$, $\boldsymbol{e}_2 = \psi^{5+}_{xz}$,
$\boldsymbol{e}_3 =\psi^{5+}_{xy}$, $\boldsymbol{B} = ( B, B ,0 )$,
\begin{equation}
 H_\mathrm{B} \sim \frac{1}{\sqrt{2}} \left(\begin{matrix} 0 & -B_z & B_y\\ B_z & 0 & -B_x\\ -B_y & B_x & 0 \end{matrix}\right) = \frac{1}{\sqrt{2}} \left(\begin{matrix} 0 & 0 & B\\ 0 & 0 & -B\\ -B & B & 0 \end{matrix}\right)\,.                                                   
\end{equation}
The eigenvectors are
\begin{equation}
 \psi^{5+}_0 = \frac{1}{\sqrt{2}}\left(\begin{matrix} 1 \\ 1 \\ 0 \end{matrix}\right)\,,\quad \psi^{5+}_{\pm 1} = \frac{1}{2}\left(\begin{matrix} -1\\1\\ \mp \mathrm{i}\sqrt{2} \end{matrix}\right)\,,
 \label{eq:MagneticFieldEigenstates}
\end{equation}
where the states can be classified according to a magnetic quantum number as given in the subscript of $\psi$ with quantization axis along the $[110]$ direction.
Note that the resulting eigenstates couple longitudinal and transversal polarizations.
They therefore allow for a SHG signal for arbitrary nonvanishing magnetic field strength if the polarizations are chosen correctly. In fact, these states can of course already be used in the degenerate case
without a magnetic field. The reason why a significant SHG signal is only visible for sufficiently high fields lies in the linewidths of the states. To the degree that the different lines overlap,
destructive interference prevents the production of SHG light.
Physical intuition for this behavior can be gained by understanding the behavior of the $\Gamma_5^+$ excitons as damped oscillations in the crystal,
\begin{equation}
 \boldsymbol{\xi}_i(t) = \boldsymbol{\xi}_i \mathrm{e}^{-\gamma t} \mathrm{e}^{\mathrm{i}\omega_i t}\,,
 \label{eq:DampedOscillation}
\end{equation}
with $i = 0,\pm 1$ denoting the oscillation modes belonging to the states $\psi^{5+}_i$ with frequencies $\omega_i = E_i/\hbar$ and
\begin{equation}
 \boldsymbol{\xi}_0 = \frac{1}{\sqrt{2}}\left(\begin{matrix} 1 \\ 1 \\ 0 \end{matrix}\right)\,, \quad \boldsymbol{\xi}_{\pm 1} = \frac{1}{2}\left(\begin{matrix} -1\\1\\ \mp \mathrm{i}\sqrt{2} \end{matrix}\right)\,.
\end{equation}
The damping $\gamma$ is proportional to the linewidths of the states.
The femtosecond pulse stimulates an initial amplitude according to
\begin{equation}
 \boldsymbol{\xi}(t=0) \sim \boldsymbol{E}^\mathrm{in} \otimes \boldsymbol{E}^\mathrm{in}\,.
\end{equation}
After the stimulation, the oscillatory modes evolve as given in Eq.~\eqref{eq:DampedOscillation}. At every time $t$, the excitonic oscillation is connected to a macroscopic polarization $\boldsymbol{P}$ via
\begin{equation}
 \boldsymbol{P}(t) \sim \boldsymbol{K} \otimes \boldsymbol{\xi} (t)\,,
\end{equation}
which will finally produce the observed SHG light at the boundary of the crystal according to $I(t) \sim |\boldsymbol{E}^\mathrm{out} \cdot \boldsymbol{P}(t)|^2$.
In the configuration considered here, the mode $\boldsymbol{\xi}_0$ does not produce a macroscopic polarization in the crystal, since $\boldsymbol{K} \otimes \boldsymbol{\xi}_0 = 0$.
The other two modes are associated with a circular polarization,
\begin{align}
 \boldsymbol{P}_{\pm 1}(t) \sim \boldsymbol{K} \otimes \boldsymbol{\xi}_{\pm 1}(t) = \pm \frac{\mathrm{i}}{2}\mathrm{e}^{-\gamma t} \mathrm{e}^{\mathrm{i}\omega_{\pm 1} t}\left(\begin{matrix} 1\\-1\\\mp\mathrm{i}\sqrt{2} \end{matrix}\right)\,.
\end{align}
Because both modes can only be stimulated through their $xy$ parts, they are excited with the same amplitude but differing sign. The total polarization
$\boldsymbol{P}_\mathrm{total}(t)$ is therefore linear with a polarization plane normal to the $[110]$ direction. The polarization vector rotates in this plane with the beat frequency $\omega_B = (\omega_{+1} - \omega_{-1})/2$
determined by the difference of the individual frequencies belonging to the oscillatory modes,
\begin{align}
\boldsymbol{P}_\mathrm{total}(t) \sim \mathrm{e}^{-\gamma t} \mathrm{e}^{i\frac{\omega_{+1} - \omega_{-1}}{2}t} \left(\begin{matrix} \cos(\omega_B t) \\-\cos(\omega_B t)\\ \sqrt{2} \sin(\omega_B t) \end{matrix}\right)\,.
\end{align}
Directly after the stimulation by the femtosecond pulse, the polarization points along the longitudinal direction $[1\overline{1}0]$ and no SHG is possible.
A SHG signal is produced to the degree that the polarization vector is rotated into the transversal $[001]$ or $[00\overline{1}]$ direction and the emitted photons are therefore polarized along the $z$ axis.
This process is determined by the competition between the Zeeman-induced beat frequency
$\omega_B$ and the damping $\gamma$. The integrated intensity and therefore the total number of detected photons is proportional to
\begin{align}
 \int I(t) \mathrm{d}t \sim \int_0^\infty \left|E^\mathrm{out}_z \mathrm{e}^{-\gamma t}\sin(\omega_B t)\right|^2 \mathrm{d}t \sim \frac{|E^\mathrm{out}_z|^2\omega_B^2}{\gamma(\omega_B^2 + \gamma^2)}\,.
 \label{eq:TotalIntensityB}
\end{align}
Since $\omega_B \sim {B}$ for small field strengths, the number of photons detected is to leading order quadratic in $\boldsymbol{B}$.

The preceding discussion reveals that only an incoming polarization exciting $\Gamma_5^+$ states transforming according to $xy$ can
produce SHG light. Returning to our goal of separating Zeeman and magneto-Stark effect, we can combine $\boldsymbol{E}^{\mathrm{out},2}~\parallel~[ 0 0 1 ]$ with $\boldsymbol{E}^{\mathrm{in},2}~\parallel~[ 1 1 0 ]$
to generate a SHG signal induced by the Zeeman effect alone.

\begin{figure*}
\includegraphics[width=0.49\textwidth]{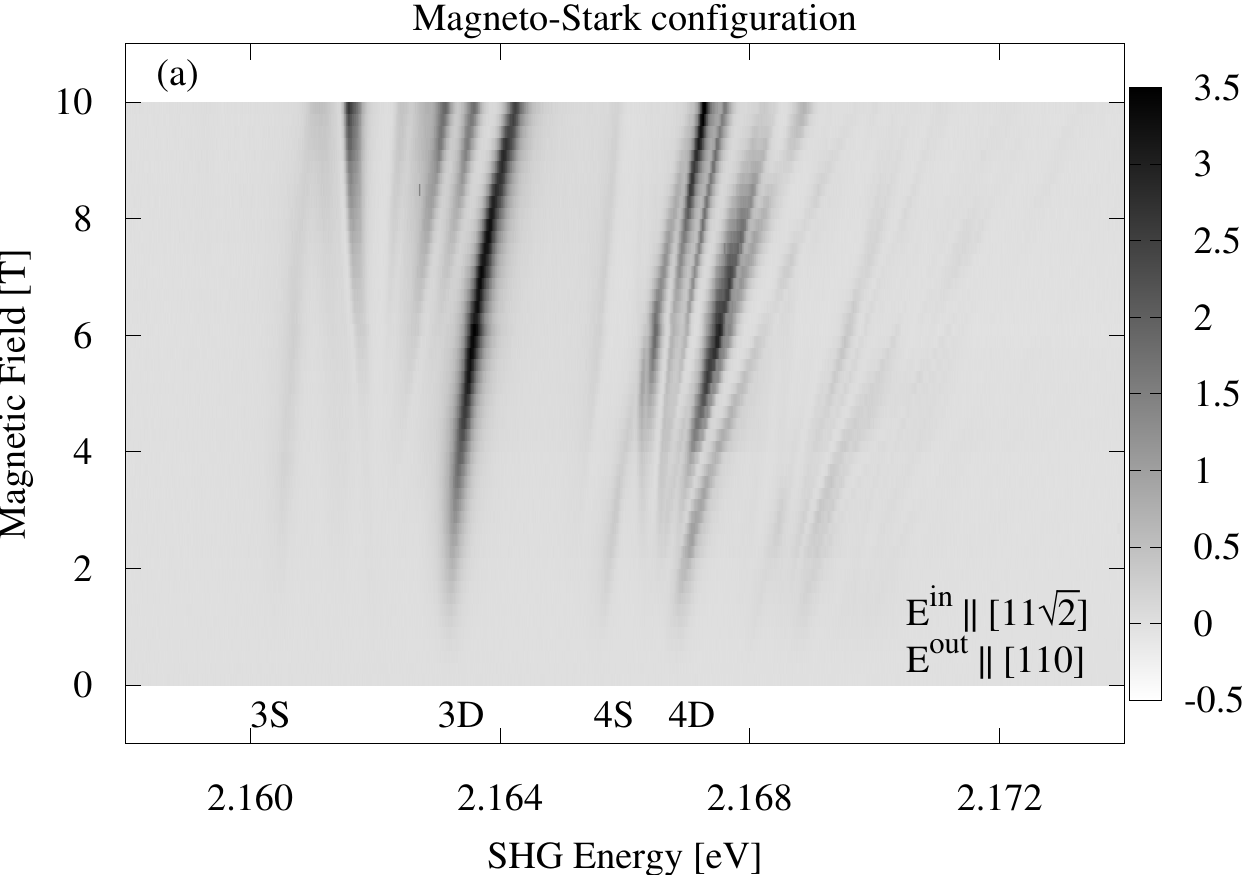} \includegraphics[width=0.49\textwidth]{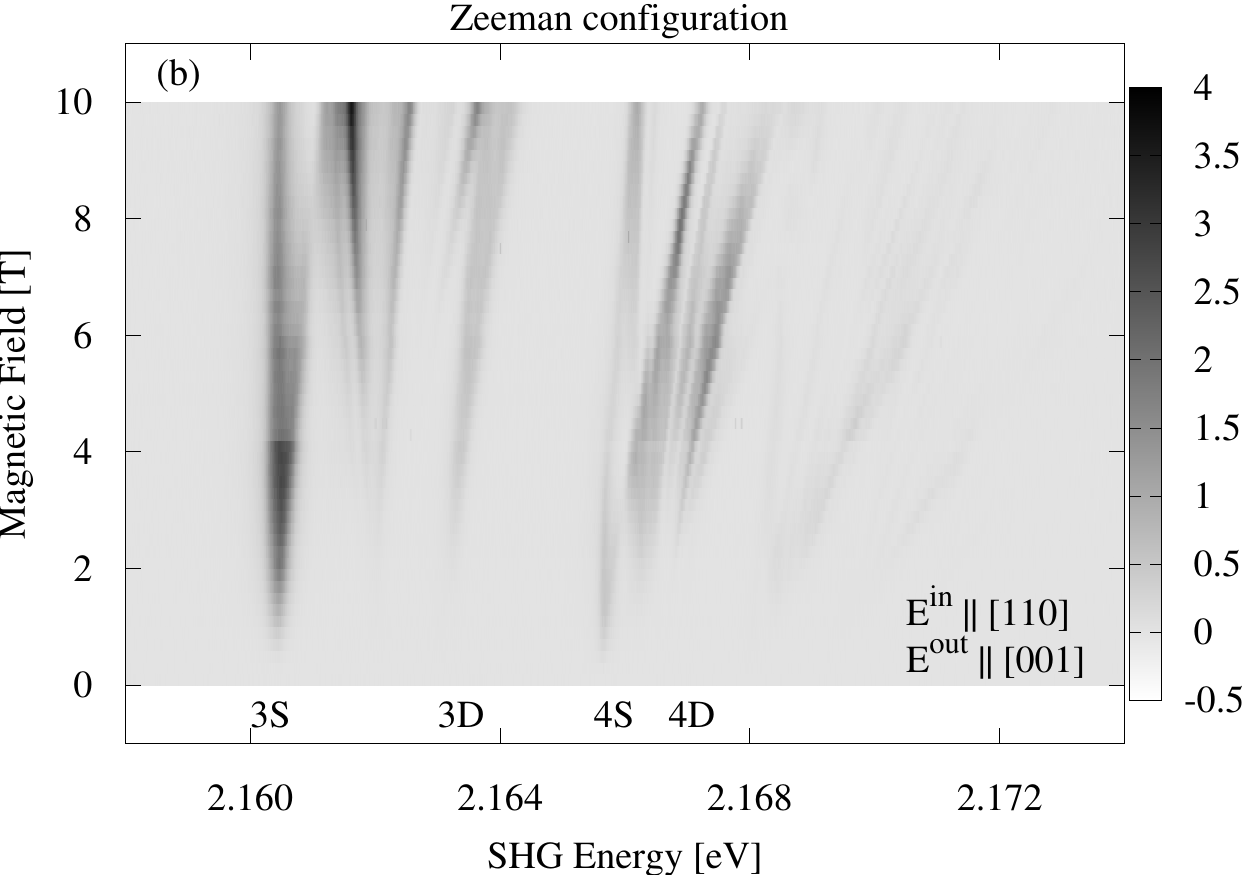}\\
\includegraphics[width=0.49\textwidth]{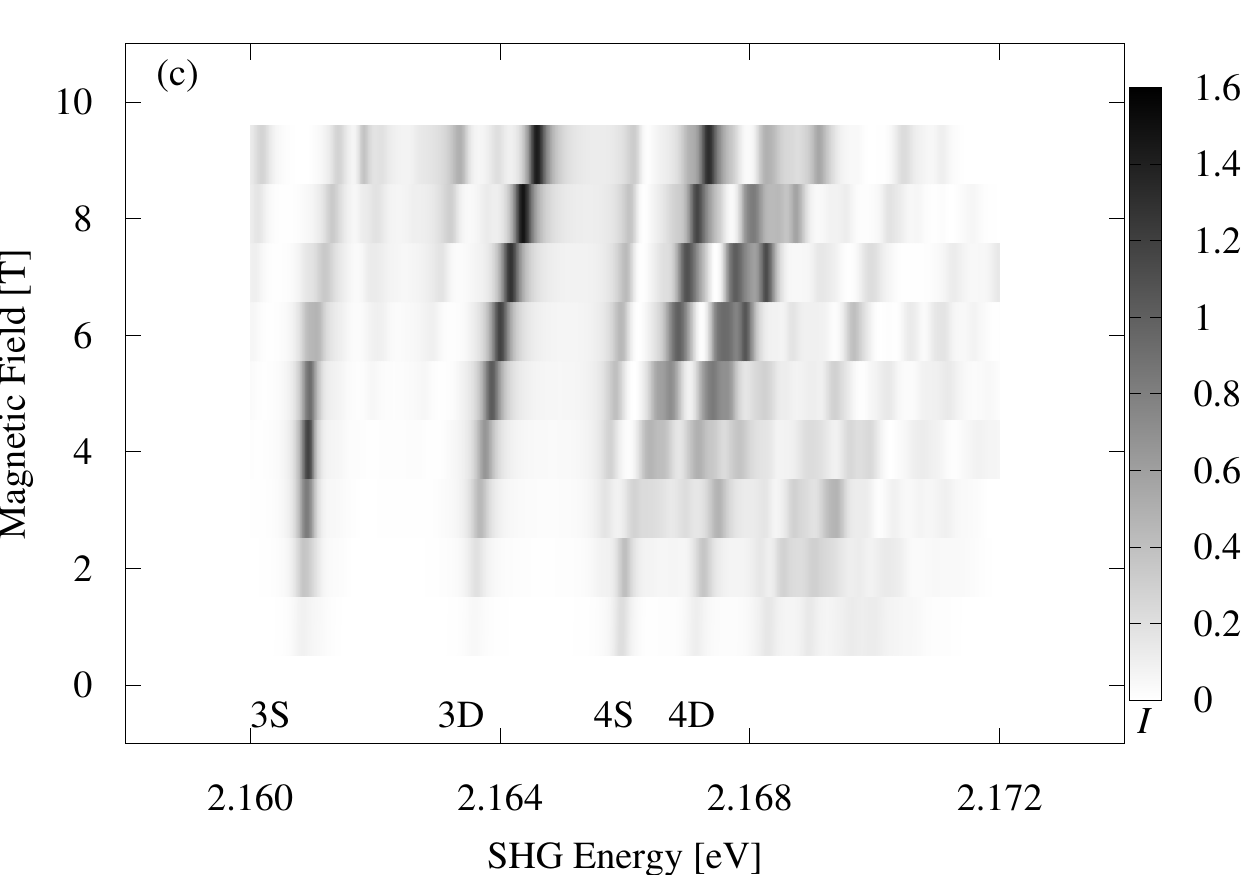} \includegraphics[width=0.49\textwidth]{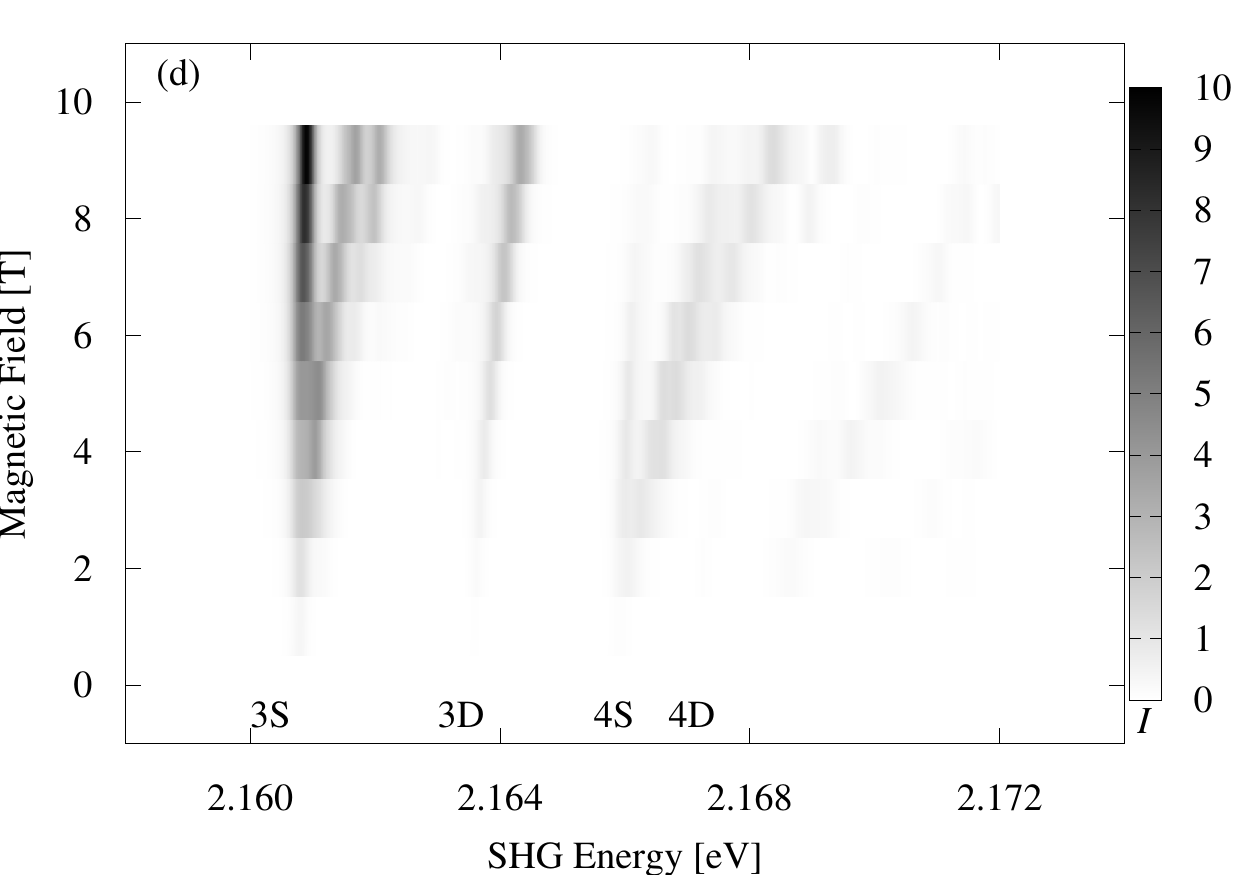} 
 \caption{Experimental SHG spectra in arbitrary units with (a) $\boldsymbol{E}^{\mathrm{in},1} \parallel \left[ 1 1 \sqrt{2} \right]$, $\boldsymbol{E}^{\mathrm{out},1} \parallel \left[ 1 1 0 \right]$ and (b)
 $\boldsymbol{E}^{\mathrm{in},2} \parallel \left[1 1 0 \right]$, $\boldsymbol{E}^{\mathrm{out},2} \parallel~\left[ 0 0 1 \right]$. The wave vector points along the $[1\overline{1}0]$ axis and the magnetic field is applied
 in $[110]$ direction. The spectra on the left hand side are mediated through the magneto-Stark effect
 and the spectra on the right are mediated by the Zeeman effect. The main features visible belong to excitons with principal quantum numbers $n=3$ and $n=4$. The corresponding numerically calculated spectra (see Eq.~\eqref{eq:SHGIntensity})
 are shown in (c) and (d).
 }
 \label{fig:Comparison_MSE_ZE}
\end{figure*}

\subsection{Additional consideration of the $\Gamma_4^-$ states}
The preceding discussion only took the $\Gamma_5^+$ states into account. We now want to consider the role of the dipole-active $\Gamma_4^-$ excitons.
To become SHG-allowed, they have to be mixed with the $\Gamma_5^+$ states. This can only happen if the inversion symmetry is broken. The magnetic field alone can therefore not induce a SHG signal mediated by odd parity states.
For this, we have to turn our attention to the magneto-Stark effect. Since the $\Gamma_4^-$ states can emit photons in a dipole process, the two-photon absorption has to be modified here to make SHG allowed.
The two-photon absorption transition amplitude for a $\Gamma_4^-$ state $\psi^{4-}_j$ due to the presence of the electric field is given by
\begin{align}
 O^{\mathcal{F},\Gamma_4^-}_{\mathrm{TPDD},j} &\sim \sum_{i,k,l} (\boldsymbol{E}^\mathrm{in} \otimes \boldsymbol{E}^\mathrm{in})_i \mathcal{F}_l \nonumber\\
             &\times \frac{\langle g | V^{DD}_i | \psi^{5+}_k \rangle \langle \psi^{5+}_k | V^\mathcal{F}_l | \psi^{4-}_j \rangle }{E_k - 2 \hbar \omega_\mathrm{in}}\,.
\end{align}
The relevant components of the two-photon operator $V^{DD}_i$ transforming as $\Gamma_5^+$ are given by
\begin{equation}
     V^{DD}_i = \sum_{j,k} \frac{|\epsilon_{ijk}|}{\sqrt{2}} \sum_l \frac{V^D_j |l\rangle \langle l | V^D_k}{E_l - \hbar \omega_\mathrm{in}}\,,
\end{equation}
with the Levi-Cevita symbol $\epsilon_{ijk}$, the dipole operators $V^D_{j,k}$ for the individual steps and the virtual intermediate states $| l \rangle$.
The components of the perturbation belonging to the electric field $V^\mathcal{F}_l$ behave as $\Gamma_4^-$.
The projection operator
\begin{equation}
  P^{5+} = \sum_k \frac{ | \psi^{4-}_k \rangle \langle \psi^{4-}_k |}{E_k - \hbar \omega_\mathrm{out}}
\end{equation}
again transforms according to $\Gamma_1^+$. The matrix elements are thus also proportional to the same coupling coefficients as in the discussion of the $\Gamma_5^+$ states.
The modified two-photon absorption amplitude is therefore given by
\begin{equation}
 \boldsymbol{O}^{\mathcal{F},\Gamma_4^-}_{\mathrm{TPDD}} = \hat{\boldsymbol{\mathcal{F}}} \otimes (\boldsymbol{E}^\mathrm{in} \otimes \boldsymbol{E}^\mathrm{in}) \,.
\end{equation}
The $\Gamma_4^-$ states emit SHG radiation by a dipole step. The SHG transition amplitude is thus proportional to
\begin{align}
 M^{\Gamma_4^-}_\mathcal{F}( \boldsymbol{E}^\mathrm{in}, \boldsymbol{E}^\mathrm{out}, \boldsymbol{\mathcal{F}}) &\sim \boldsymbol{O}_\mathrm{TPDD,\mathcal{F}}\cdot\boldsymbol{O}_\mathrm{D} \nonumber\\
                                                                                            &= (\hat{\boldsymbol{\mathcal{F}}} \otimes (\boldsymbol{E}^\mathrm{in} \otimes \boldsymbol{E}^\mathrm{in}))\cdot \boldsymbol{E}^\mathrm{out}\nonumber\\
                                                                                            &= (\boldsymbol{E}^\mathrm{in} \otimes \boldsymbol{E}^\mathrm{in}) \cdot (\hat{\boldsymbol{\mathcal{F}}} \otimes \boldsymbol{E}^\mathrm{out})\,,
\end{align}
and we get the same formula as for the $\Gamma_5^+$ states. Our conclusions regarding the polarization dependencies of the Zeeman-induced and magneto-Stark-induced SHG amplitudes thus remain unchanged by the additional consideration of the
$\Gamma_4^-$ excitons. In particular, it remains the case that with the combination of polarizations given by $\boldsymbol{E}^{\mathrm{in},1}$ $\parallel$ $[ 1 1 \sqrt{2} ]$ and $\boldsymbol{E}^{\mathrm{out},1}$ $\parallel$ $[ 1 1 0 ]$
only SHG induced by the magneto-Stark effect is visible and with the combination of polarizations given by $\boldsymbol{E}^{\mathrm{in},2}$ $\parallel$ $[ 1 1 0 ]$ and $\boldsymbol{E}^{\mathrm{out},2}$~$\parallel$~$[ 0 0 1 ]$ only SHG
induced by the Zeeman effect is visible.

These combinations of polarizations for the incoming and outgoing light allow for the separation of the Zeeman and magneto-Stark effect to the degree that the approximations made in the
preceding discussion are valid. In the first configuration with $\boldsymbol{E}^{\mathrm{in},1}$ and $\boldsymbol{E}^{\mathrm{out},1}$, quadrupole emission is forbidden entirely. 
Restricting our treatment to the dominant contributions, only the electric-field induced mixture of $\Gamma_4^-$ and $\Gamma_5^+$ excitons can produce any
SHG signal at all, even for strong fields. For the second configuration with $\boldsymbol{E}^{\mathrm{in},2}$ and $\boldsymbol{E}^{\mathrm{out},2}$, only weaker statements are possible. The electric-field induced SHG vanishes
only if the Zeeman splitting between the states is small.
Still, if the energetic distance of a SHG-active $\Gamma_5^+$ multiplet to the dipole-active $\Gamma_4^-$ states is large, the contribution of dipole emission processes remains minor.
This effect can be seen in Fig.~\ref{fig:DetermineA}, where the second combination of polarizations is used. The high energy 3D line shows an especially
small influence of the electric field, its intensity being almost unaffected by variations in the strength of dipole emission processes. This is probably explained by its high energetic distance to the 3P lines and other odd parity states
as stated above.\\
Apart from allowing for the separation of Zeeman and magneto-Stark effect, the formulas for the SHG amplitudes derived in this section can be used for the detailed discussion of the polarization dependencies of the SHG signal.
Since $M_\mathcal{F}( \boldsymbol{E}^\mathrm{in}, \boldsymbol{E}^\mathrm{out}, \boldsymbol{\mathcal{F}})$ and the amplitude induced by the magnetic field are different functions of the polarizations,
the effects can be distinguished experimentally. Complementary to the discussion here, this is done in the manuscript by A. Farenbruch \emph{et al.}~\cite{Farenbruch2019}, where the polarization dependencies
for SHG processes other than the ones considered here are studied as well.

\begin{figure}
\includegraphics[width=\columnwidth]{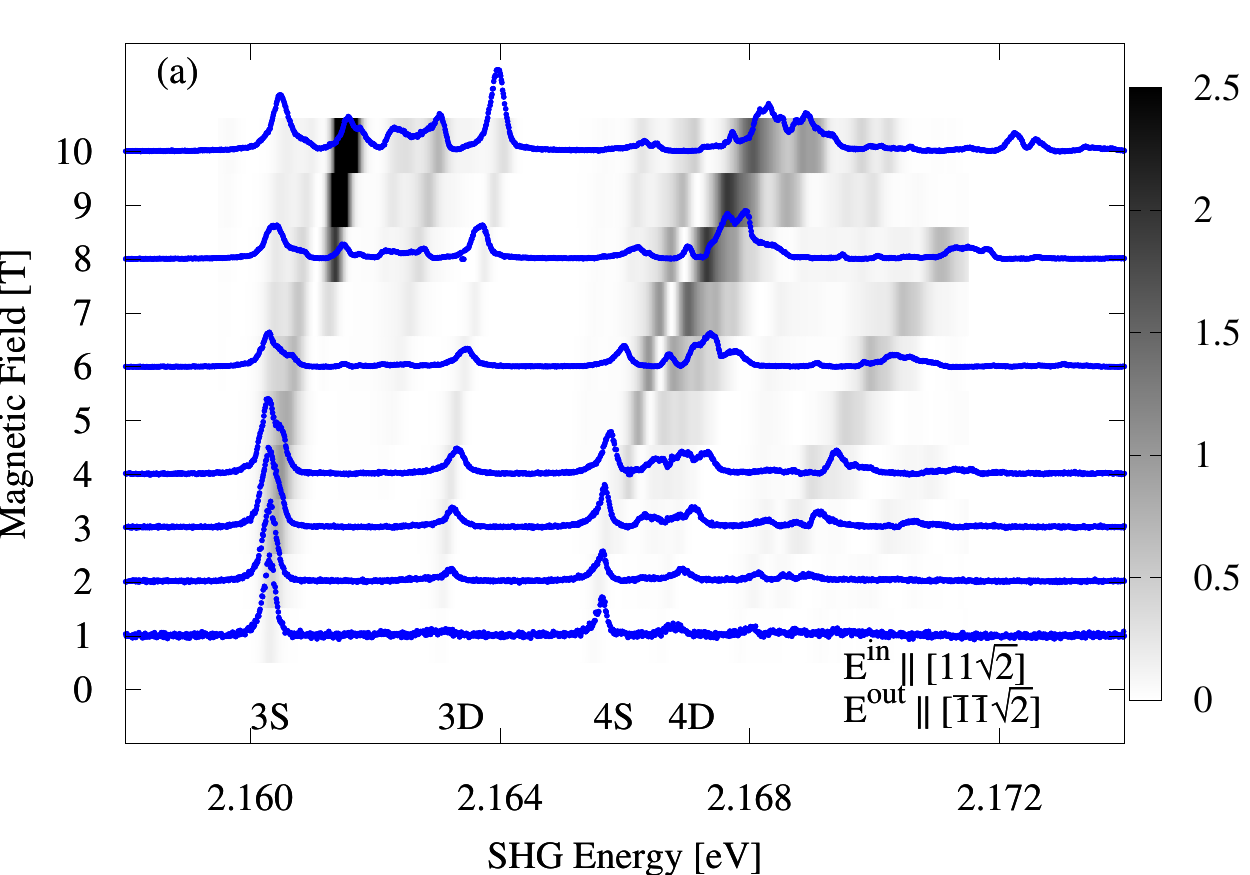}\\
\includegraphics[width=\columnwidth]{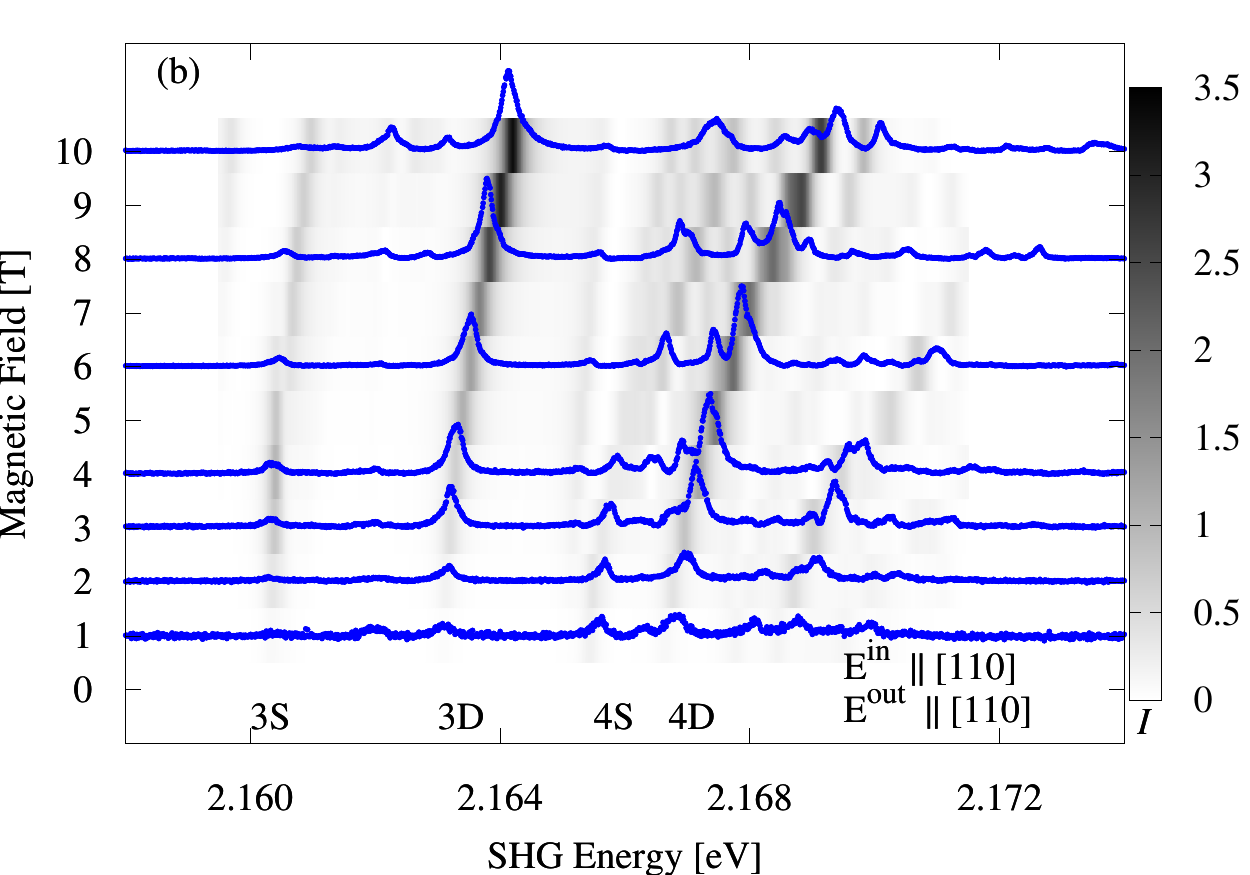}
\caption{Experimental SHG spectra (blue {lines}) with $\boldsymbol{B}$~$||$~$[001]$, $\boldsymbol{K}$~$||$~$[1\overline{1}0]$ and (a) $\boldsymbol{E}^\mathrm{in}$~$||$~$[11 \sqrt{2}]$, $\boldsymbol{E}^\mathrm{out}$~$||$~$[\overline{1}\overline{1} \sqrt{2}]$ and 
(b) $\boldsymbol{E}^\mathrm{in}$~$||$~$[110]$, $\boldsymbol{E}^\mathrm{out}$~$||$~$[110]$. The corresponding numerically simulated spectra {(grayscale) have been shifted by $-0.5\,\mathrm{meV}$ to allow for a better comparison.} The main visible features belong to excitons with principal quantum numbers $n=3$ and $4$. The feature visible at $E \approx 2.162 \, \mathrm{eV}$, $B \approx 8 - 10 \, \mathrm{T}$ in the numerical spectrum in panel (a) has an intensity exceeding
the color palette scale and is most likely due to some numerical artifact.\\
}
\label{fig:VoigtForbiddenAdditional}
\end{figure}
\begin{figure}
\includegraphics[width=\columnwidth]{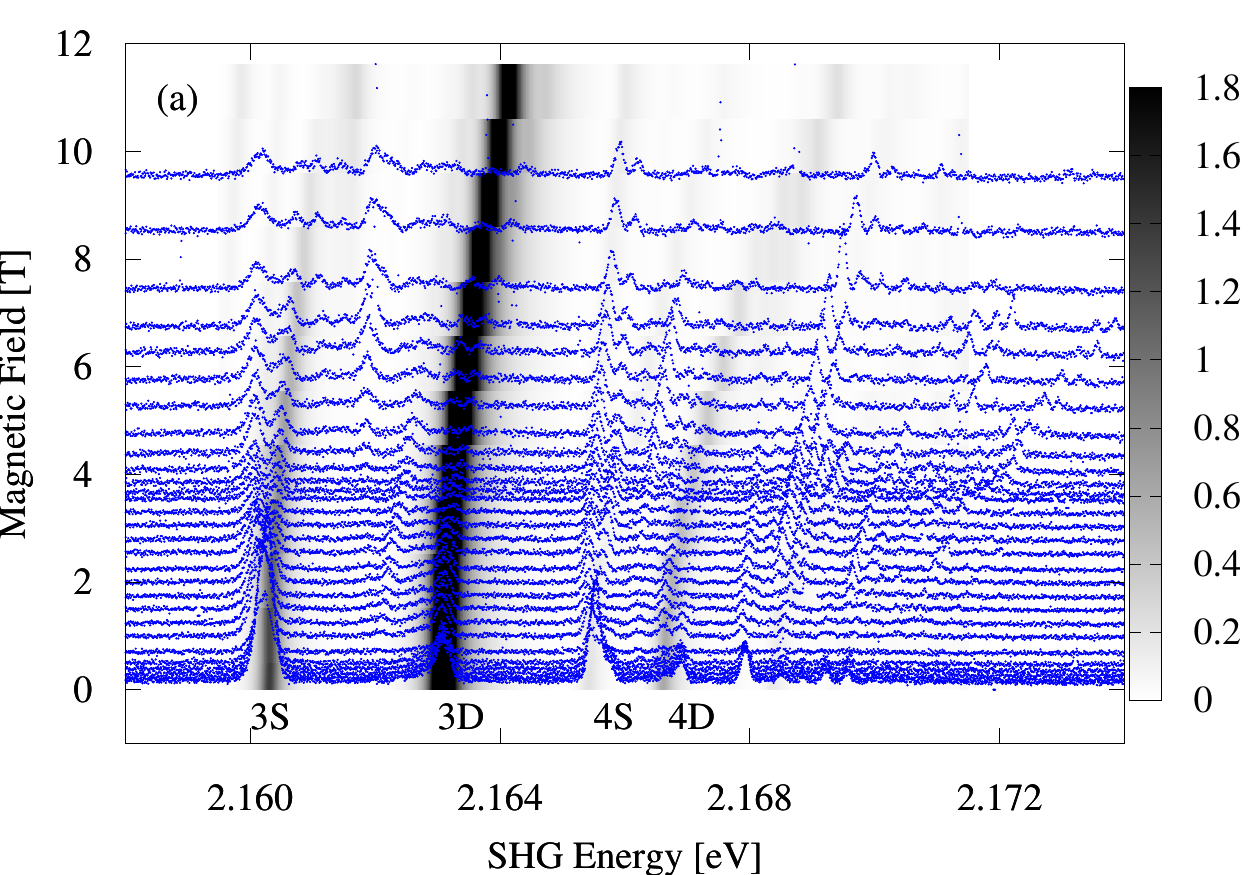}
\includegraphics[width=\columnwidth]{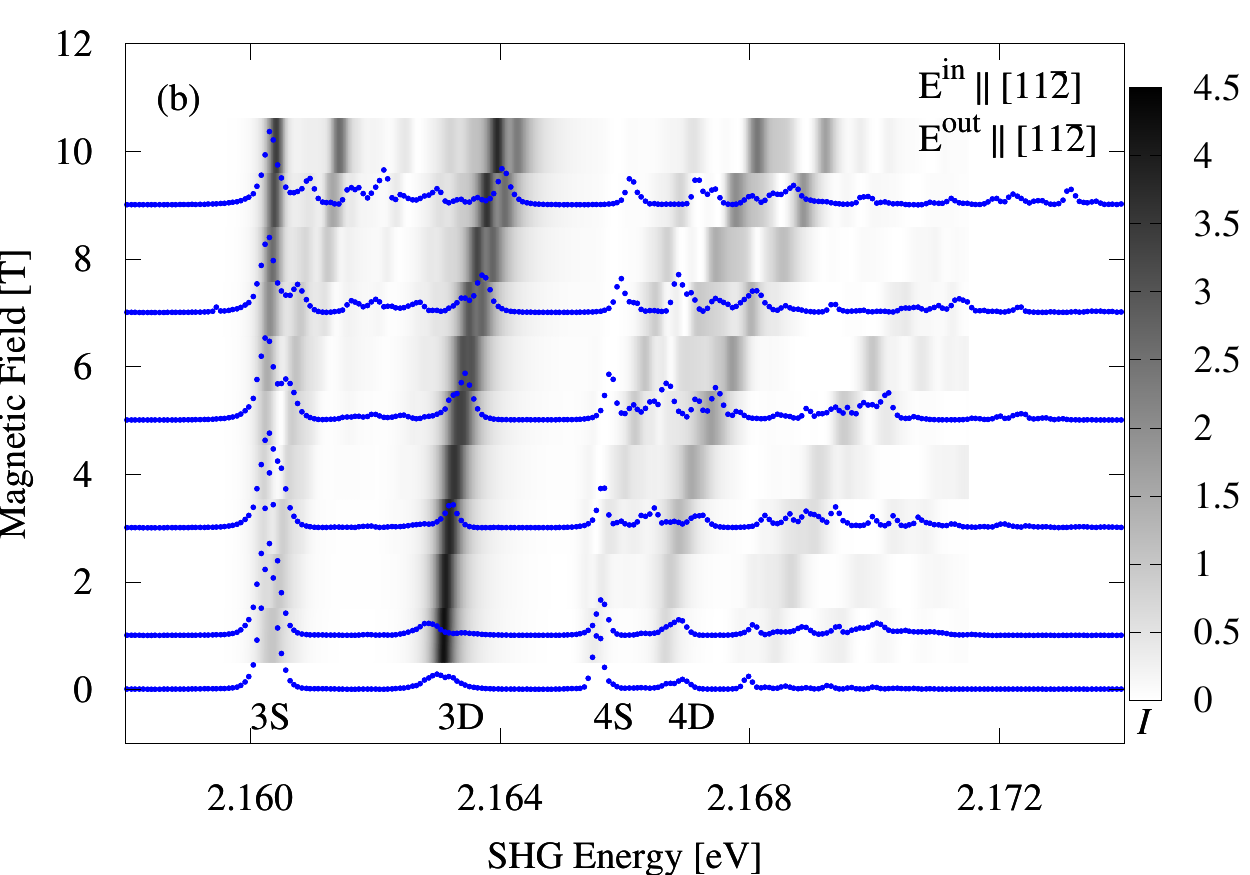}
 \caption{Experimental spectra {(blue lines)} with $\boldsymbol{K} \parallel [111]$,  $\boldsymbol{E}^\mathrm{in} \parallel \boldsymbol{E}^{\mathrm{out}} \parallel [11\overline{2}]$ in
 (a) Faraday configuration with $\boldsymbol{B} \parallel [111]$ and (b) Voigt configuration with $\boldsymbol{B} \parallel [1\overline{1}0]$.
 {The corresponding numerically simulated spectra (shifted by $-0.5\,\mathrm{meV}$) are shown in grayscale.}
 The main features visible belong to excitons with principal quantum numbers $n=3$ and $4$. Note that the 3D line in panel (a) exceeds the upper limit of the gray scale.~\label{fig:FaradayVoigtAllowed}}
\end{figure}

\section{Experiment}
\label{sec:Experiment}
Details of the experimental set-up are reported in Ref.~\cite{Mund2018}. By replacing the $0.5\,\mathrm{m}$ monochromator by a $1\,\mathrm{m}$ monochromator we improved the spectral resolution
from $100\,\mu\mathrm{eV}$ to $16\,\mu\mathrm{eV}$. Details of the new set-up are shown in the complementary publication by Farenbruch \emph{et al.}~\cite{Farenbruch2019}. The Cu$_2$O samples are cut from a natural crystal in different crystalline
orientations and thicknesses. They are mounted strain-free in a split-coil superconducting magnet allowing a magnetic field strength of up to $10\,\mathrm{T}$ in Faraday and Voigt configuration. With the use of
half-wave plates the linear polarization of the incoming (laser) light and outgoing (SHG) light can be varied continuously and independently. The measurements were taken in superfluid helium at about $1.4\,\mathrm{K}$.
For excitation we used a tunable femtosecond laser ($200\,\mathrm{fs}$, spectral width of $10\,\mathrm{meV}$).
The frequency-doubled intensity profile is centered on $2.164\,\mathrm{eV}$ with a FWHM of $14\,\mathrm{meV}$, cf.\ Ref.~\cite{Mund2018}.
To take its influence into account for the numerical calculations, we weight the numerically obtained spectrum with a Gaussian function with the appropriate parameters.

\section{Results and Comparison with Experiment}
\label{sec:Results}

In Fig.~\ref{fig:Comparison_MSE_ZE}, both experimental and numerical spectra with the polarizations discussed in Sec.~\ref{sec:SelectionRules} are shown. A general agreement between experiment and numerical spectra is observed. Some discrepancies remain:
For both spectra, the numerical features in the region of the 3S states are too strong. This is probably due to the central-cell corrections as explained at the end of Sec.~\ref{subsec:RelStrength}.

In general, the SHG spectrum is determined by a combination of the Zeeman and magneto-Stark effect. In Fig.~\ref{fig:VoigtForbiddenAdditional}, we show additional examples of magnetic-field-induced
SHG spectra in a forbidden direction. The used combination of polarizer and analyzer in Fig.~\ref{fig:VoigtForbiddenAdditional} (a) produces a spectrum that is a product of both the Zeeman and magneto-Stark effect in full generality,
whereas Fig.~\ref{fig:VoigtForbiddenAdditional} (b) shows another spectrum entirely produced through the MSE, since $\boldsymbol{K} \otimes \boldsymbol{E}^\mathrm{out} = 0$ in this case.
Here too, in both cases reasonable agreement between experiment and numerical simulation is achieved. In the numerical data in Fig.~\ref{fig:VoigtForbiddenAdditional} (a), a strong feature appears for
$E \approx 2.162 \, \mathrm{eV}$, $B \approx 8 - 10 \, \mathrm{T}$ that is not seen in the experiment. The two remarks from the end of section~\ref{subsec:RelStrength} apply here: The inaccuracies in the central-cell corrections and the linewidths
lead to an overestimated SHG intensity. Fig.~\ref{fig:VoigtForbiddenAdditional} (b) on the other hand shows generally good agreement.

In Fig.~\ref{fig:FaradayVoigtAllowed}, we show pictures of SHG along the allowed direction $[111]$.
Some agreement is observed, but there are also significant differences. Most evidently, the D excitons are stronger than the S excitons in the numerical data, but in the experiment the reverse is the case. A possible
explanation is to be found in the treatment of the center-of-mass motion. Due to the inversion symmetry of cuprous oxide, the SHG signal in the field-free case can be thought of as being induced by the finite wave vector $\boldsymbol{K}$.
For $\boldsymbol{B} \neq 0$, this will give an additional contribution to the spectrum that requires a more careful consideration of the center-of-mass motion than the one used here. For SHG in forbidden directions, the center-of-mass
motion by itself does not induce a SHG signal, and thus our treatment is sufficient in that case.
As expected, the Voigt configuration as seen in Fig.~\ref{fig:FaradayVoigtAllowed} (b) shows more features compared to the Faraday configuration in Fig.~\ref{fig:FaradayVoigtAllowed} (a) due to the additional mixing caused by the electric field.

\section{Conclusion}
\label{sec:Conclusion}
We extended the method developed by Schweiner \emph{et al}.\ for the calculation of absorption spectra of excitons in Cu$_2$O~\cite{ImpactValence,frankmagnetoexcitonscuprousoxide,frankevenexcitonseries} to the simulation
of Second Harmonic Generation intensities.

In Cu$_2$O, SHG is forbidden along axes with a C$_2$ symmetry. The application of an external magnetic field makes SHG along those directions allowed.
In this paper, we mainly consider the case of 
SHG along forbidden axes. We identify two separate mechanisms by which a magnetic field can induce a SHG signal. First of all, the magnetic field itself reduces the symmetry
and mixes the exciton states in an appropriate way to produce a nonvanishing SHG intensity. In this case, parity remains a good quantum number and the emitted photon can only be produced by a quadrupole process. In the Voigt configuration, the magnetic field
induces an additional effective electric field. This breaks the inversion symmetry and also makes SHG with dipole emission processes possible.

We study spectra where both quadrupole and dipole emission processes play a role. To this end, we estimate the relative strength of these by comparing suitable numerical and experimental
spectra.

We compare numerically calculated and experimental data for various choices of polarizations of the incoming and outgoing light, direction of the wave vector and direction of the magnetic field.
We find that for certain configurations, spectra are to leading order entirely induced by the
magnetic or by the electric field. Good agreement between experiment and theory is observed for the most part, some weaknesses of the numerical method remain.

First of all, the treatment of SHG in allowed directions will require a more careful approach towards the center-of-mass motion, since in this case, the nonvanishing $\boldsymbol{K}$ vector will
by itself induce a SHG signal. To include this properly, the Hamiltonian will have to be complemented by additional $\boldsymbol{K}$-dependent terms.

The SHG intensities associated with specific exciton lines is dependent on their linewidths. The inclusion of this effect in our model is only rudimentary. A better treatment is difficult, since it would require
the detailed knowledge of the lifetimes of the exciton states even in the regime of strong mixing.

An additional weakness of the numerical method used here are the central-cell corrections. Due to their inaccuracy, the positions of the even exciton states are slightly too high energetically. This leads to a too strong
mixing of the S and P states and thus to too strong intensities of these lines. An improved treatment of the central-cell corrections could solve this problem.

Differences between theory and experiment may arise also from imperfections in experiment: there might be slight misalignment of the crystal relative to the targeted geometry,
which is, however, not expected to exceed a few degrees. Also the polarization selection might not be perfect with an accuracy of about one degree.
In Ref.~\cite{Mund2019} it was shown that SHG is sensitive to strain down to levels of ppm. Therefore, also strain may influence the appearance of the spectra.

{Still, for the main application considered in this work, that is, for the investigation of magnetic-field induced SHG spectra in forbidden directions, we achieve satisfactory results. Improved treatments of the 
central-cell corrections and center-of-mass motion in allowed 
configurations are necessary in future work.}

\acknowledgments
The theoretical studies at University of Stuttgart were supported by Deutsche Forschungsgemeinschaft
(Grant No.\ MA1639/13-1). The experimental studies at TU Dortmund were supported by the Deutsche Forschungsgemeinschaft through the International Collaborative Research Centre TRR 160 (Projects No. C8 and A8)
and the Collaborative Research Centre TRR 142 (Project No. B01).
We also acknowledge the support by the project AS  459/1-3. We thank Frank Schweiner for his contributions.

%

\end{document}